\input harvmac
\input epsf
\sequentialequations
\def\figin#1{#1}
\def\ifig#1#2#3{\xdef#1{fig.~\the\figno}
\goodbreak
\midinsert\figin{\centerline{#3}}
\smallskip\centerline{\vbox{\baselineskip12pt
\advance\hsize by -1truein\noindent\footnotefont{\bf Figure.~\the\figno:} #2}}
\endinsert\global\advance\figno by1}

\def\ads{A_{dS}}
\def\arn{A_{RN} }

\def\rshell{r_{s}(t) }
\def\gammamax{{\gamma_{max}}^2 }
\def\alphao{\alpha_{0} }

\def\nin{{n_{dS}}^{r} }
\def\nout{{n_{RN}}^{r} }

\def\vds{V_{dS} }
\def\vrn{V_{RN} }
\def\zin{z_{dS} }
\def\zina{z_{dS-} }
\def\zinb{z_{dS+} }

\def\zouta{z_{RN-} }
\def\zoutb{z_{RN+} }
\def\zmax{z_{max} }
\def\zmin{z_{min} }
\def\zst{z_{st} }
\def\R+{R_{+} }
\def\Rext{R_{ext} }
\def\I{\uppercase\expandafter{\romannumeral1} }
\def\II{\uppercase\expandafter{\romannumeral2} }
\def\III{\uppercase\expandafter{\romannumeral3} }
\def\IV{\uppercase\expandafter{\romannumeral4} }
\def\V{\uppercase\expandafter{\romannumeral5} }
\def\VI{\uppercase\expandafter{\romannumeral6} }
\def\threem{\uppercase\expandafter{\romannumeral3}_{-}}
\lref\bardeen{J. Bardeen, Proceedings of the GR5 Meeting, Tiflis, U.S.S.R. (1968).}
\lref\mars{M. Mars, M.M. Martin-Prats and J.M.M. Senovilla, Class. Quant. Grav. 13 (1996) L51.}
\lref\bordetwo{A. Borde, Phys. Rev. D50 (1994) 3392.}
\lref\bordethesis{A. Borde, {\it Singularities in Classical Spacetimes}, Ph.D. dissertation, 
SUNY - Stony Brook (1982).}
\lref\bais{F.A. Bais and R.J. Russell, {\it Magnetic Monopole Solution of Nonableian Gauge 
Theory in Curved Spacetime}, Phys. Rev. D11 (1975) 2692, erratum Phys. Rev. D12 (1975) 3368.}
\lref\freund{Y.M. Cho and P.G.O. Freund, {\it Gravitating `T Hooft Monopoles}, 
Phys. Rev. D12 (1975) 1588, erratum Phys. Rev. D13 (1976) 531.}
\lref\linde{A. Linde and D. Linde, {\it Topological Defects as Seeds for Eternal Inflation}, 
Phys. Rev. D50 (2456) 1994, e-print hep-th/9402115.}
\lref\vilenkin{A. Vilenkin, {\it Topological Inflation}, Phys. Rev. Lett. 72 (1994) 3137,
e-print hep-th/9402085.}
\lref\penrose{R. Penrose, {\it Gravitational Collapse and Spacetime Singularities}, 
Phys. Rev. Lett. 14 (1965) 57.}
\lref\penhawk{S.W. Hawking and R. Penrose, {\it The Singularities of Gravitational Collapse and Cosmology}, Proc. Roy. Soc. Lond. A314 (1970) 529.}
\lref\borde{A.Borde, {\it Regular Black Holes and Topology Change}, 
Phys. Rev. D55 (1997) 7615, e-print gr-qc/9612057} 
\lref\maeda{T. Tachizawa, K. Maeda, T. Torii, {\it Nonabelian Black Holes and Catastrophe 
Theory 2: Charged Type}, 
Phys. Rev. D51 (1995) 4054, e-print gr-qc/9410016 }
\lref\van{P. Van Nieuwenhuizen, D. Wilkinson and M.J. Perry, 
{\it On a Regular Solution of t'Hooft's Magnetic Monopole in Curved Space},
Phys. Rev. D13 (1976) 778}
\lref\lnw {K. Lee, V.P. Nair, E.J. Weinberg, {\it Black Holes in Magnetic Monopoles}, 
Phys. Rev. D45 (1992) 2751, e-print hep-th/9112008 }
\lref\sakai{N. Sakai, {\it Dynamics of Gravitating Magnetic Monopoles}, 
Phys. Rev. D54 (1996) 1548, e-print gr-qc/9512045.}
\lref\kt{D. Kastor, J. Traschen, {\it Horizons Inside Classical Lumps},
Phys. Rev. D46 (1992) 5399, e-print hep-th/9207070}
\lref\boulware{D. Boulware, 
{\it Naked Singularities, Thin Shells and the Reissner-Nordstrom Metric}, 
Phys. Rev. D8 (1973) 2363.}
\lref\eardley{D. Eardley and S.J. Kolitch, {\it Quantum Decay of Domain Walls in Cosmology: 
1. Instanton Approach}, Phys. Rev. D56 (1997) 4651, e-print gr-qc/9706011. }
\lref\bgg{S. Blau, E. Guendelman, A. Guth, {\it The Dynamics of False Vacuum Bubbles}, 
Phys. Rev. D35 (1987) 1747.}
\lref\cho{I. Cho, J. Guven, {\it Modelling the Dynamics of Global Monopoles},
Phys. Rev. D58 (1998) 063502, e-print gr-qc/9801061}
\lref\cheng{T. Cheng and L. Li, {\it Gauge Theory of Elementary Particle Physics}, 
Oxford University Press (1984).}
\lref\israel{W. Israel, {\it Singular Hypersurfaces and Thin shells in General 
Relativity}, Nuovo Cimento 44B (1966) 1; erratum Nuovo Cimento 48B (1967) 463}
\lref\lowe{G.L. Alberghi, D. Lowe and M. Trodden, {\it Charged False Vacuum Bubbles and the AdS/CFT Correspondence}, 
JHEP 9907:020 (1999), e-print  hep-th/9906047.}
\lref\guven{G. Arreaga, I. Cho and J. Guven, {\it Stability of Self-Gravitating Monopoles}, 
e-print gr-qc/0001078.}


\Title{\vbox{\baselineskip12pt
\hbox{hep-th/0002220 }}}
{\vbox{\centerline{\titlerm The Dynamics of Collapsing Monopoles}
\bigskip\centerline{and Regular Black Holes}
 }}
\bigskip
\centerline{Hyunji Cho, David Kastor and Jennie Traschen}
\medskip
\centerline{\it Department of Physics and Astronomy}
\centerline{\it University of Massachusetts}
\centerline{\it Amherst, MA 01003-4525 USA}
\medskip\bigskip
\centerline{\bf Abstract}
\medskip
We study the formation and stability of regular black holes by 
employing a thin shell approximation to the dynamics of collapsing magnetic monopoles.
The core deSitter region of the monopole is matched across the shell to a Reissner-Nordstrom 
exterior.  We find static configurations which are nonsingular black holes
and also oscillatory trajectories about these
static points that share the same causal structure.
In these spacetimes the shell is always hidden behind the black hole horizon.
We also find shell trajectories that pass through the asymptotically flat region and 
model collapse of a monopole to form a regular black hole.
In addition there are trajectories in which the deSitter core encompasses a 
deSitter horizon and hence undergoes topological inflation. 
However, these always yield singular black holes and never have the shell
passing through the aymptotically flat region.
Although the regular black hole spacetimes satisfy the strong energy condition, 
they avoid the singularity theorems by failing to satisfy the genericity 
condition on the Riemann tensor. 
The regular black holes undergo a change in spatial topology in accordance with 
a theorem of Borde's.

\bigskip
\medskip
\Date{February, 2000}
\vfill\eject
\newsec{Introduction}

The purpose of this paper is to investigate the formation and stability of regular
black holes. In 1968 Bardeen \bardeen\ presented an example of a black
hole spacetime that satisfies the weak energy condition, contains a region
of trapped surfaces, and yet has no curvature singularities.
Recently Borde \borde\  proved a theorem that helps to clarify when regular black holes
can occur. Borde showed that if a black hole spacetime contains trapped surfaces, 
satisfies the weak energy condition and is non-singular, 
then there must be a change of topology in the spacetime. Inside the
horizon there is a region where the topology changes from  open to
compact spatial slices. 
The false-vacuum cores of topological defects typically
satisfy the weak energy condition and have internal geometry that is approximately
described by deSitter spacetime.  Further, static magnetic monopole black
hole spacetimes have been found in which the horizon is embedded in the monopole
fields \lnw  , rather than being completely collapsed into the singularity \bais\freund. 
Therefore, monopole spacetimes are good candidates for examples of
regular black holes, illustrating Borde's theorem in a physically interesting 
context.  

We have found two further illustrative examples of regular black holes
in the literature, though neither was commented on as such by the authors.
Boulware's paper on dynamical charged dust shells \boulware\ contains a
spacetime diagram of a charged shell which collapses to form
a regular $Q=M$ extremal black hole. The singularity is completely covered by the
Minkowski interior of the shell. Tachizawa et. al. \maeda\  
display three spacetime diagrams, reproduced in figures (1a,c,d) below that 
follow from approximate solution for a static, gravitating monopole \lnw .
In these figures the exterior, unshaded region of the 
spacetime is magnetically charged Reissner-Nordstrom (RN) and the interior, 
shaded region is deSitter (dS), which approximates the monopole core. 
Depending on the values for the magnetic charge and the cosmological
constant, these spacetimes have either (a) $Q>M$ and no horizons, 
(c) $Q<M$ with black hole horizons, but no deSitter horizon, or 
(d) $Q<M$ with black hole and deSitter
horizons.  There is also an extremal $Q=M$ case shown in figure (1b).
The three black hole spacetimes are regular.
The spatial slices $S_1$ and $S_2$ show the transition from open $S^2\times R$ 
to compact $S^3$ spatial sections.
\ifig\fone{Static monopole shell configurations. The dashed lines denote the static monopole shell. 
The unshaded regions are Reissner-Nordstrom.  The shaded regions are deSitter.
The corresponding ranges of the dimensionless parameter $\eta$ are (a) $\eta<3/2$, 
(b) $\eta=3/2$, (c) $3/2<\eta\sqrt{3}$ and (d) $\sqrt{3}<\eta$.  }
{\epsfysize=1.75in \epsfbox{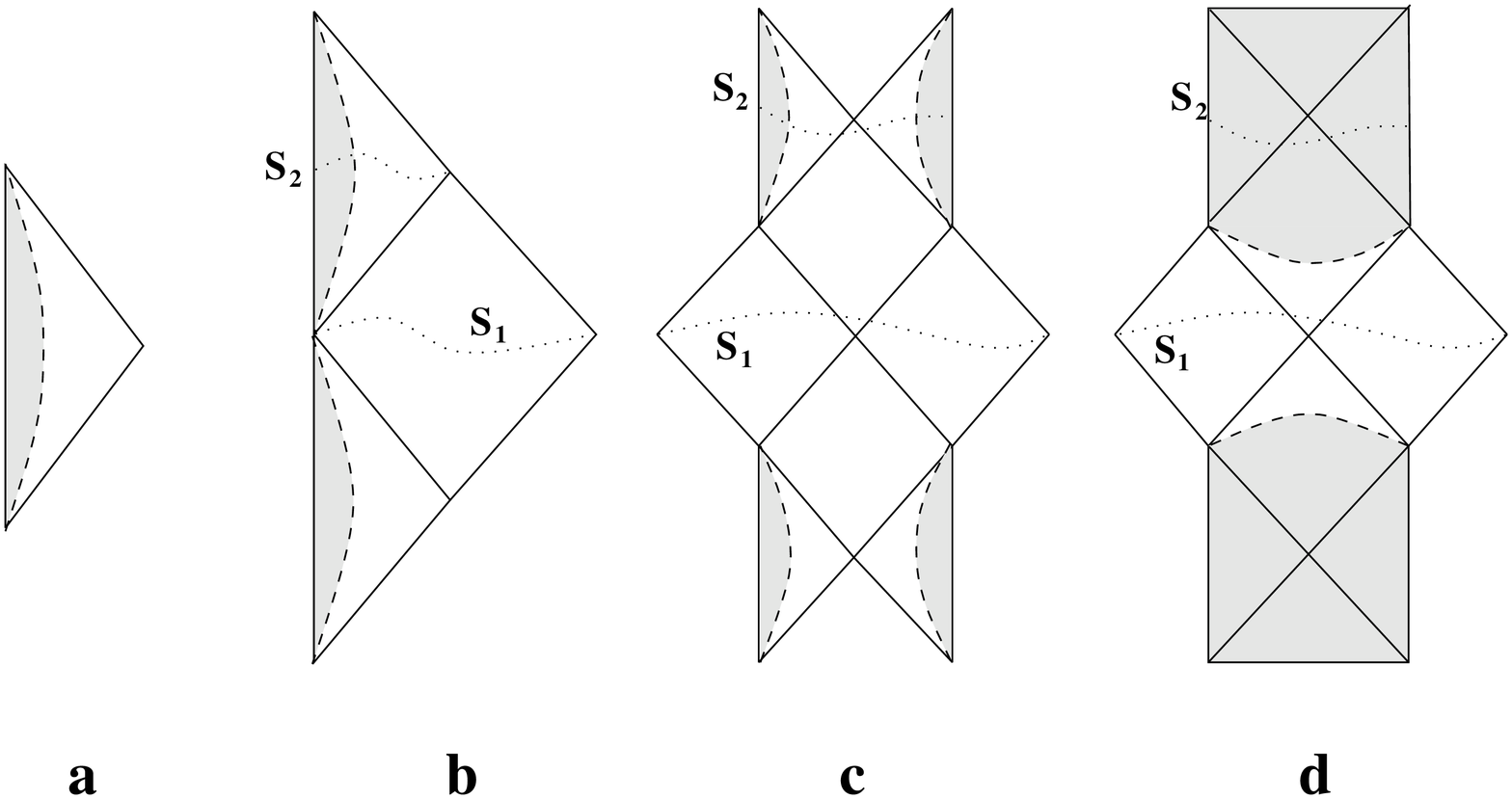}}\vskip -8pt

We would like to study the formation of such regular black holes via collapse.
The dynamics of the full Einstein-YM-Higgs system has been studied
numerically in \sakai.  In order to obtain analytic results, here we will make use 
of a thin shell approximation for the monopole instead. The system will be described by the false
vacuum energy density, or cosmological constant $\Lambda$, the mass $M$ of
the spacetime, the charge $Q$ and the mass per unit area $\sigma$ of the shell\foot{
After this work was completed, references \lowe\ and \guven\
appeared which also study the dynamics of magnetic monopoles using the thin shell approximation 
employed here.  These papers include many of the same results, 
although with somewhat different emphasis in the analysis.  
In particular, reference \lowe\ includes the possibility of a cosmological constant in the region exterior
to the monopole and uses the AdS/CFT correspondence to address the uniqueness of evolution of the shell trajectory 
through Cauchy horizons.  Reference \guven\ gives a complete analysis of possible trajectories for the monopole shell in terms of the mass $M$ and charge $Q$ of the spacetime and the ratio $k$ of the interior false vacuum energy density to the surface energy density of the shell.}.
Clearly one looses the detailed non-abelian structure
monopole in this model.  At the end of the introduction we will argue
that the  shell approximation is qualitatively correct when the energy density
of the monopole core is dominated by the false vacuum energy.

Before adding shells of stress energy, consider the following static, 
spherically symmetric model of a gravitating monopole \lnw\maeda,
\eqn\metric{ds^2 = -A (r) dt^2 + {dr^2 \over A (r)} +r^2 d\Omega ^2.
}
The core energy density
is dominated by the vacuum potential energy, and far from the core
the energy density goes to zero like $r^{-4}$.  The spacetime should
interpolate between a core deSitter region and a magnetically charged Reissner-Nordstrom exterior. 
In fact, the two regions can be matched directly across a charged shell at radius $R$ giving
\eqn\static{\eqalign{
&A(r)=\ads(r)  = 1- H^2 r^2 \   ,\  \  r< R\cr
&A(r)=\arn(r)=  1-{2M\over r} +{ Q^2 \over r^2 }   ,\  \    r>R  \cr } }
Requiring
that the metric and its first derivative be continuous fixes the 
matching radius $R$ and
the ADM mass $M$ in terms of the charge $Q$ and the cosmological constant
$\Lambda=3H^2$, 
\eqn\match{
{R\over Q}={1\over \eta},\qquad
{M\over Q}={2\over 3}\eta ,\qquad \eta^2=\sqrt{3}QH.
}
References \lnw\maeda\
discuss the metric \static\  as an approximation to their more detailed numerical 
results.  We want to stress here that
the metric \static\  subject to \match\  is actually a solution to
the Einstein-Maxwell equations. There is {\it no} shell of stress energy
at the matching radius, but there is a shell of charge.  Futher it is interesting
to note that if one wants to match Minkowski to Schwarzchild \eardley   ,
 Minkowski to RN \boulware  , or deSitter to
Schwarzchild \bgg  , a shell of stress energy is needed to do the
matching.  A deSitter interior can also be matched directly to flat space
minus a solid angle, the asymptotic exterior of a global
monopole with no shell of stress energy. Reference \cho\ then further adds a shell to study
the dynamics of this model of a global monopole.
 
The resulting spacetime diagrams from the matching
are then those of figure (1), with the different cases
corresponding to different ranges of the dimensionless variable $\eta$.
For $0<\eta < 3/2$ we have $Q>M$ and there are no horizons
in the spacetime as in figure (1a). 
Therefore, this can be considered a weakly gravitating monopole. 
For $\eta=3/2$ we have $Q=M$ as in figure (1b).  For $3/2<\eta<\sqrt{3}$ we have $Q<M$ with the
matching radius $R$ such that both black hole horizons are present outside the shell, but no deSitter
horizon inside.  For $\sqrt{3}<\eta$ we have 
$Q<M$ with only the outer black hole horizon present outside the matching radius.  
In this case the inner deSitter region extends through a cosmological horizon as well.
Note that in this case the
boundary between regions is a spacelike surface, and therefore is more like a phase transition
in the spacetime. We will not be considering spacetimes of this type in the present work.

When one adds  a shell of stress energy at the matching radius, the
spacetime becomes dynamical.  The shell generically collapses, expands,
or oscillates. The spacetimes described by  \static\
are robust in the sense that
there continue to be static shell solutions, and also solutions 
in which the shell oscillates about the static points. In these cases
the black holes are regular, since the shell radius is bounded
away from zero and inside the shell the
spacetime is always deSitter. 
There are regular extremal and non-extremal black holes of this type. 
In the extended RN spacetime there is a sequence
of repeating timelike singularities as one moves up the diagram.
In order for the extended spacetime to be regular, one would need to cover each
singularity with a shell. From the asymptotically flat region these
black holes look just like any RN black hole.

A second type of dynamical solution describes a monopole that collapses to form a black hole.
These are shells that pass through the aymptotically flat region 
for some portion of the trajectory.  One can imagine starting the evolution
of a monopole as the shell passes through this region and watching it collapse 
inside its horizon. It turns
out that all shells that pass through the asymptotically flat region
correspond to shell oscillations around a local mininum of an effective
potential. 
In the extended spacetimes, these
oscillations take the shell repeatedly out through white hole horizons and back into black hole horizons.
Since the shell radius is bounded, the interior of
the shell is regular. In particular, in the extremal case a single oscillating
shell, which passes through the asymptotically flat region, 
covers up all the singularities of the RN spacetime, giving a maximally
extended regular black hole spacetime.
In the non-extremal case, however,
there are RN singularities on both sides of the diagram, only one set of which
is covered up by a single oscillating shell trajectory.
One would have to add additonal shells to cover up the singularity on the other
side of the spacetime.

The possibility of topological inflation has been studied in the literature \linde\vilenkin.
This refers to the case when the core of a topological defect is approximately described by
a deSitter metric and includes a deSitter, or cosmological, horizon.
There are many shell trajectories in the present case which contain
inflating cores. All of these are singular black holes, and the shell
never passes through the asymptotically flat region in any of the inflating
cases. So this is a kind of null result.  In this stripped down model,
an observer in the asymptotically flat region would never see a monopole
collapse which is bound to evolve into an inflating cosmology.

How do the regular black hole spacetimes we find in this paper, 
as well as the earlier Bardeen example, 
avoid forming singularities? Penrose's (1965)
singularity theorem \penrose\  requires that the spacetime be globally
hyberbolic, and the later (1970) singularity theorem of Hawking and
Penrose \penhawk\ requires instead a genericity condition on the Riemann tensor.
None of these example are generic, or globally hyperbolic. The 
regular $Q<M$ examples illustrate the spatial topology change required
by Borde's theorem.  He discusses how, although light rays are focused
in the region of trapped surfaces, the change to the three-sphere topolgy 
allows focusing without the formation of  a singularity. The extremal, $Q=M$
spacetimes do not contain trapped surfaces and therefore do not fall within the
scope of Borde's theorem.  Nonetheless, the spatial topology changes inside the 
horizon.

We conclude the introduction with a simple comparision of the
shell model for gravitating monopoles to features of nongravitating monopoles.
Compare the relations \match  ,
which give the mass and core radius of the static gravitating monopole
spacetime in terms of the charge and vacuum
energy density, to analogous relations for non-gravitating monopoles. 
An `t Hooft-Polyakov monopole is described by three
parameters; the Higgs vacuum expectation value $v$, its self-coupling
$\lambda $ and the gauge coupling constant $e$ (see {\it e.g.} \cheng). 
The monopole then has charge $Q\sim 1/e$, mass $M\sim v/e$ and core radius $R\sim 1/(ev)$.
This implies an average energy density of the core $\rho \sim e^2 v^4$.
Now, define a `Hubble constant' for the monopole by
$H^2 \sim \rho \sim e^2 v^4$. Trading the parameters $Q,H$ for $v,e$
gives $M\sim  ( Q^3 H )^{1/2}$
 and $R\sim \left( {Q \over  H} \right)^{1/2}$, which
have the same functional dependence on $Q,H$ as in \match  .

Where did the parameter $\lambda$ go to? For the core to be dominated
by the vacuum energy density, and hence for the metric in the core region to
be well approximated by
deSitter, one would expect $H^2\sim V_{false} \sim \lambda v^4 $.
In order that this  expression for $H$ agree with our previous estimate
in the monopole core, we need that $e^2 \sim \lambda$. So for such
monopoles, we have  this qualitative motivation that the simple
gravitational model \metric  ,\static  , \match\ is a model of
a monopole.

\newsec{Collapsing Magnetic Monopoles}

%
We model a collapsing magnetic monopole 
by a charged, spherically symmetric domain wall, or shell, that separates a region of deSitter
spacetime inside the shell from a region of Reissner-Nordstrom spacetime outside. 
The domain wall has constant surface energy 
density $\sigma$, total magnetic charge $Q$ and time dependent radius $R(\tau)$, where
$\tau$ is the proper time for a worldline with constant angular position on the domain wall.
The spacetime metric then has the form \metric\ with
\eqn\scndmetric{A (r) =\left\{
\eqalign{&\ads(r),\qquad  r< R(\tau) \cr
&\arn(r), \qquad r> R(\tau)\cr}\right .}
By Gauss's law the total charge $Q$ of the shell must be the same as the charge
$Q$ appearing in the exterior RN metric function \static.

\vskip 0.1in\noindent
{\it Shell Dynamics}\vskip 0.05in

The dynamics of the shell are determined using the Israel shell formalism \israel , which 
imposes Einstein's equation including
the distributional stress-energy of the shell. In the present case, there are also contributions 
to $T_{ab}$ from
the cosmological constant inside the shell and the Maxwell stress-energy outside.
The motion of the shell is described by its four-velocity, which we take to be
radially directed, $u^a =u^t {\partial\over \partial t}+u^r{\partial \over \partial r}$,
and normalized, $u^a u_a =-1$. The unit normal to the shell is also radially directed,
and satisfies $n^a u_a =0$, $n^a n_a = +1$. In general the stress-energy of a spherically symmetric
shell is described by two parameters, the mass per unit area $\sigma$ and the pressure $p$.
Israel's shell formalism relates the jump in the extrinsic curvature $K_{ab}$ of the shell 
to integrals of the shell stress-energy. The jump in $K_{\tau\tau}$ gives
the evolution equation for $\sigma$, $\dot \sigma =-2{\dot R \over R}(\sigma -p )$.
Dust shells have zero pressure, and so  $\sigma \sim R^{-2}$ which
implies that the total mass of the shell is constant. Domain walls have $\sigma =p$
and so $\sigma$ is constant.  In this work we model the monopole shell as a domain wall.

The other jump condition can be summarized simply in  terms
of the jump in the radial components of the outward pointing 
unit normal vectors to the shell, $n^r_{dS}$ and $n^r_{RN}$,
which are evaluated just inside and just outside the shell respectively,
\eqn\boundary{
r[K^\theta _\theta ]^{dS}_{RN} =  n^r_{dS} - n^r_{RN}  =4 \pi \sigma R(\tau).} 
For $Q=0$ this reduces to the case studied in \bgg.
This condition determines the motion of the shell as follows.
The squares of the radial normal components satisfy
\eqn\squares{(n^r_{dS})^2=A_{dS}(R) + \dot R^2,\qquad (n^r_{RN})^2=A_{RN}(R) + \dot R^2,}
where $\dot R=dR/d\tau$.  
Squaring equation \boundary\ and substituting in \squares\ yields the relations
\eqn\firstorder{n^r_{RN} = {1\over 8\pi\sigma R}(A_{dS}
-A_{RN}-16\pi^2\sigma^2 R^2),\qquad
n^r_{dS}={1\over 8\pi\sigma R}(A_{dS}
-A_{RN}+16\pi^2\sigma^2 R^2)}
Squaring these equations and using \squares\ again then gives the alternate forms
\eqn\motion{\eqalign{
\dot R^2&={1\over (8\pi\sigma R)^2}(A_{dS}-A_{RN}-16\pi^2\sigma^2 R^2)^2-A_{RN}\cr
\dot R^2&={1\over (8\pi\sigma R)^2}(A_{dS}-A_{RN}+16\pi^2\sigma^2 R^2)^2-A_{dS}\cr}}
Following reference \bgg, we introduce rescaled dimensionless coordinates and variables.
The rescaled radial coordinate of the shell $z(\tau^\prime)$ is regarded as a function of the rescaled
time parameter $\tau^\prime$ with
\eqn\rescale{
z = \left (H^2 + 16 {\pi}^2 {\sigma}^2 \over 2M \right)^{1/3}R,\qquad
\tau' = \left({H^2 + 16 {\pi}^2 {\sigma}^2} \over {8 \pi \sigma} \right)
\tau }
We then rewrite equation \motion\ in the form of particle motion in an effective potential $V(z)$.
\eqn\eomz
{\left(dz \over d\tau' \right)^2 + V (z) = E} 
with the potential given in terms of dimensionless parameters $\alpha$, $\gamma$ by
\eqn\vz{V (z) = -{1 \over z^2} {\left(z^2 - {1 \over z} + {\alpha \over z^2}
\right)}^2 - {\gamma}^2 \left({1 \over z} - {\alpha \over z^2} \right).}
The energy $E$ and the parameters $\alpha$, $\gamma$ are given by
\eqn\variables{E=-{64\pi^2\sigma^2\over (2Mh^2)^{2/3}},\qquad
\alpha={Q^2 h^{2/3}\over (2M)^{4/3}},\qquad
\gamma={8\pi\sigma\over h},}
where $h^2=H^2+16\pi^2\sigma^2$. Note that $\gamma$ varies only over the range $0\le\gamma\le 2$.
The problem began with four dimensional parameters $H$, $M$, $Q$ and $\sigma$.  Through
the rescaling, this has been reduced to the three dimensionless parameters $\alpha$, $\gamma$
and $E$.  

\vskip 0.1in \noindent
{\bf\it Shape of the Potential}
\vskip 0.05in
One now analyzes the different types shell motion  by
studying the possible shapes for the potential depending on the parameters $\alpha$ and $\gamma$ 
and then considering different values of the energy $E$ for a fixed potential.
The potential $V(z)\rightarrow -\infty$ both as $z\rightarrow 0$ and as $z\rightarrow\infty$.
In between, the potential may, or may not, have a local minimum depending on $\alpha$, $\gamma$ (see
figures (2a) and (2b)).
For $\alpha=0$, which gives the $Q=0$ case studied in \bgg, there is never a
local minimum.  The repulsive Coulomb self-interaction of the $\alpha>0$ shell is therefore responsible
for the possibility of stable static configurations as in figure (2b).  Such local minima are also present
in the analysis of global monopoles in \cho.
\ifig\fone{The potential $V(z)$ can either (a) have no local minimum, or (b) have a local minimum,
depending on the values of the parameters $\alpha$ and $\gamma$. In (2a) we have sketched in the 
deSitter horizon curve $E(z_{dS})$ and shown a shell trajectory that crosses the horizon. Figure (2b)
includes the Reissner-Nordstrom horizon curve and trajectories with energues 
$E_A$ that falls below this curve and hence
has $Q>M$ and $E_B$ that has $Q<M$ and crosses both inner and outer RN horizons.}
{\epsfysize=1.6in \epsfbox{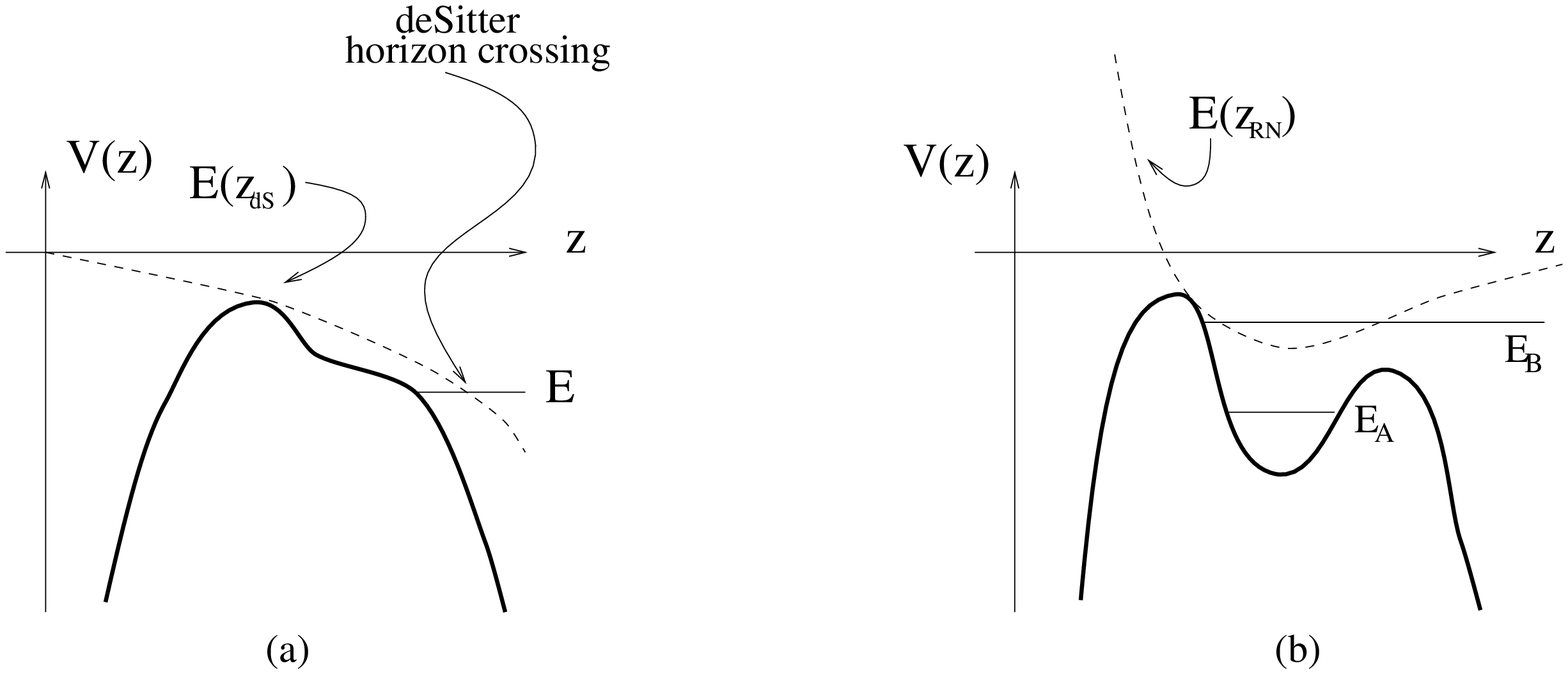}}\vskip -8pt\noindent

In detail, we find that the potential $V(z)$ always has a local minimum if 
$0<\alpha < \alpha_1$, with $\alpha_1=\left( 1 \over 2\right) ^ {4/3}$. For $\alpha$ in the range
$ \alpha_1< \alpha < \alpha_2$, with $\alpha_2={3 \over {4^{4/3}}}$, 
there exists a value ${\gamma_{max}}^2 $ that depends on $\alpha$, 
such that $V(z)$ has a local minimum only for 
${\gamma}^2 < {\gamma_{max}}^2$. 
If $\alpha_2<\alpha$, then $V(z)$ does not have a local minimum for any value of $\gamma$.

\vskip 0.1in \noindent
{\bf\it Horizon Curves}
\vskip 0.05in
Given fixed values for the parameters $H$, $M$, $Q$ and $\sigma$, the deSitter and 
Reissner-Nordstrom metrics respectively have horizons at
\eqn\rhorizons{r_{dS}={1\over H},\qquad r_{RN}^{\pm}=M\pm\sqrt{M^2-Q^2}.}
The corresponding dimensionless coordinates for the horizons,
\eqn\horizons{
z_{dS} = {{2 \sqrt {|E|}} \over {\gamma (4 -{\gamma}^2)^{1/2}}},\qquad 
z_{RN}^\pm = {{\gamma}^2 \pm \sqrt { {\gamma}^4 - 4|E| \alpha {\gamma}^2} \over 
{2|E|}}}
are then
functions of the parameters $\alpha$ and $\gamma$ that determine the shape of the potential
and also of the particle energy $E$.  
Whether, or not, a shell trajectory crosses through the various horizon
radii determines the causal structure of the resulting spacetime.  
Since the horizon radii expressed in terms of the dimensionless radial variable 
$z$ are energy dependent, it is useful to plot
horizon curves on the potential diagrams as well.  
In order to do this, we invert the relationships \horizons\ to get the 
energies $E_{dS}$ and $E_{RN}$ that correspond to given values of the horizon radii
\eqn\horizonenergies{
E_{dS}(z_{dS})= {-{{\gamma}^2 \left( 1 - {{\gamma}^2 \over 4 } \right) z_{dS}^2}},\qquad
E_{RN} (z_{RN})= 
{-{{\gamma}^2 \left({ 1 \over z_{RN}} - {\alpha \over (z_{RN})^2}\right)}} }
%
A shell trajectory then passes through a given horizon if its constant energy line on the
potential diagram intersects the corresponding horizon curve. For example, 
in figure (2a) we have sketched in
both the deSitter horizon curve and the trajectory of a shell that passes through the deSitter horizon.

From equation \horizons\ we can see that
in terms of the dimensionless variables, horizons exist only for $4\alpha |E|\le \gamma^2$, with
$4\alpha |E|= \gamma^2$ being the extremal $Q=M$ limit.  The Reissner-Nordstrom 
horizon curve has the limit $E_{RN}(z)\rightarrow\infty$
for $z\rightarrow 0$ and approaches zero from below as $z\rightarrow\infty$.  In between it has a minimum
at $E_{ext}=-\gamma^2/4\alpha$.  Shell trajectories with $E>E_{ext}$ correspond to $Q<M$ spacetimes, 
trajectories with $E=E_{ext}$ have $Q=M$ and trajectories with $E<E_{ext}$ have $Q>M$.

\vskip 0.1in\noindent
{\bf\it Signs of the radial normals}
\vskip 0.05in
One can see from equation \firstorder\ that the radial normals $n_{dS}^r$ and $n_{RN}^r$ can have
either sign and that in particular $n_{RN}^r<0$ for sufficiently large, or small, shell radius $R$. 
By definition the normal vectors point away from the deSitter region and into the Reissner-Nordstrom
region.  The sign of $n_{RN}^r$ determines whether, in moving into the Reissner-Nordstrom region the
radial coordinate is increasing, or decreasing.  

The actual signs of $n_{dS}^r$ and $n_{RN}^r$ are very useful in analyzing the motion.
We can see that the points where these quantities pass through zero are given by the places
where the horizon curves $E_{dS}(z)$ and $E_{RN}(z)$ intersect the potential $V(z)$.
First rewrite the expressions \firstorder\ for $n_{dS}^r$ and $n_{RN}^r$ in terms of 
dimensionless quantities
\eqn\norminz{
n^r_{dS} =  {-\left(1 - {
\gamma^2 \over 2 } \right) z^4 + z - \alpha\over z^3\sqrt{-E}},\qquad
n^r_{RN} = {-z^4 +z - \alpha \over z^3\sqrt {-E}}
}
Now rewrite the horizon curves $E_{dS}(z)$ and $E_{RN}(z)$ separating out explicity
a factor of the potential $V(z)$
\eqn\hrznlns{\eqalign{
E_{dS}(z) &=
 V (z) +{\left \{ {1 \over z^3} {\left[{- \left( 1 - \gamma^2 
\over 2 \right)} z^4 + z - \alpha \right]} \right \}}^2\cr
E_{RN}(z) &=  V (z) + 
{ \left[{-z^4 +z - \alpha\over z^3 } \right]}^2 \cr
}}
We see that the radial normals $n_{dS}^r$ and $n_{RN}^r$ 
vanish precisely when the horizon curves the potential meet. 

The radial normals will have zeroes only for over certain ranges of the
parameters $\alpha$ and $\gamma$. In detail we find that
for $\alpha < \alpha_2$, 
the potential $V(z)$ and the Reissner-Nordstr\"om horizon line meet at two radii, 
$\zouta$ and $\zoutb$. Within the range $\zouta<z<\zoutb$, 
the radial normal $n_{RN}^r>0$, while for $z<\zouta$ and $z>\zoutb$ the radial normal 
$\nout<0$. For $\alpha > \alpha_2$, the horizon line is always above the potential and 
$\nout$ is always negative.

For the sign of the radial normal $\nin$\ we find 
that for ${\gamma}^2 \ge 2$, the potential and the De Sitter horizon
curve meet each other at a single radius $\zin$. 
The radial normal satisfies $\nin <0$ for $z < \zin$ and $\nin \ge 0$ for $z\ge \zin$ independent
of the value of $\alpha$. 
For ${\gamma}^2 < 2$, the sigrn on $\nin$ depends 
on $\alpha$.  For $\alpha < \alpha_2$ and $\gamma<2$, the potential and the deSitter 
horizon curve meet at two points, which we label $\zina$ and $\zinb$. For 
$\zina \le z \le \zinb$, we have $\nin \ge 0$, with $\nin \le 0$ otherwise. 
When $\alpha_2<\alpha $ and $ 2\left( 1 - {\alpha_2^3\over\alpha^3}\right) < {\gamma}^2 < 2$, 
the result is the same. 
However, for $ {\gamma}^2 <2\left( 1 - {\alpha_2^3\over\alpha^3}\right)$, 
$\nin<0$ always holds..

\newsec{Summary of Results}

Clearly, there are many possible trajectories for the shells.  
In this section we will give a summary of the different types of motion and the resulting 
causal structures.  Our primary interest is in whether, or not, there are shell trajectories 
in which the  monopole collapses to form a regular black hole.
A full tabulation of different types of shell motions is given in appendix A.  
In particular, table (1) in appendix A lists the different possible types of shell motion viewed from 
the Reissner-Nordstrom side of the spacetime.  In figures (10)-(16) examples of these trajectories are
shown on potential diagrams that include the RN horizon line and the points $z_{RN\pm}$ where the 
radial normal $\nin$ changes sign.  Table (2) and figures (17)-(19) give similar information with 
respect to the deSitter portion of the spacetime.  Finally, table (3) relates the forms of potentials 
and horizon lines in figures (10)-(19) to specific ranges of the parameters $\alpha$ and $\gamma$.
It may be helpful for the reader to refer to these figures while reading through the present section.

In the figure below, we reproduce figure (2b) with different
shell energies drawn in for reference.
\ifig\fone{Sketch of a potential diagrams showing shell energy levels of different basic types.}
{\epsfysize=1.75in \epsfbox{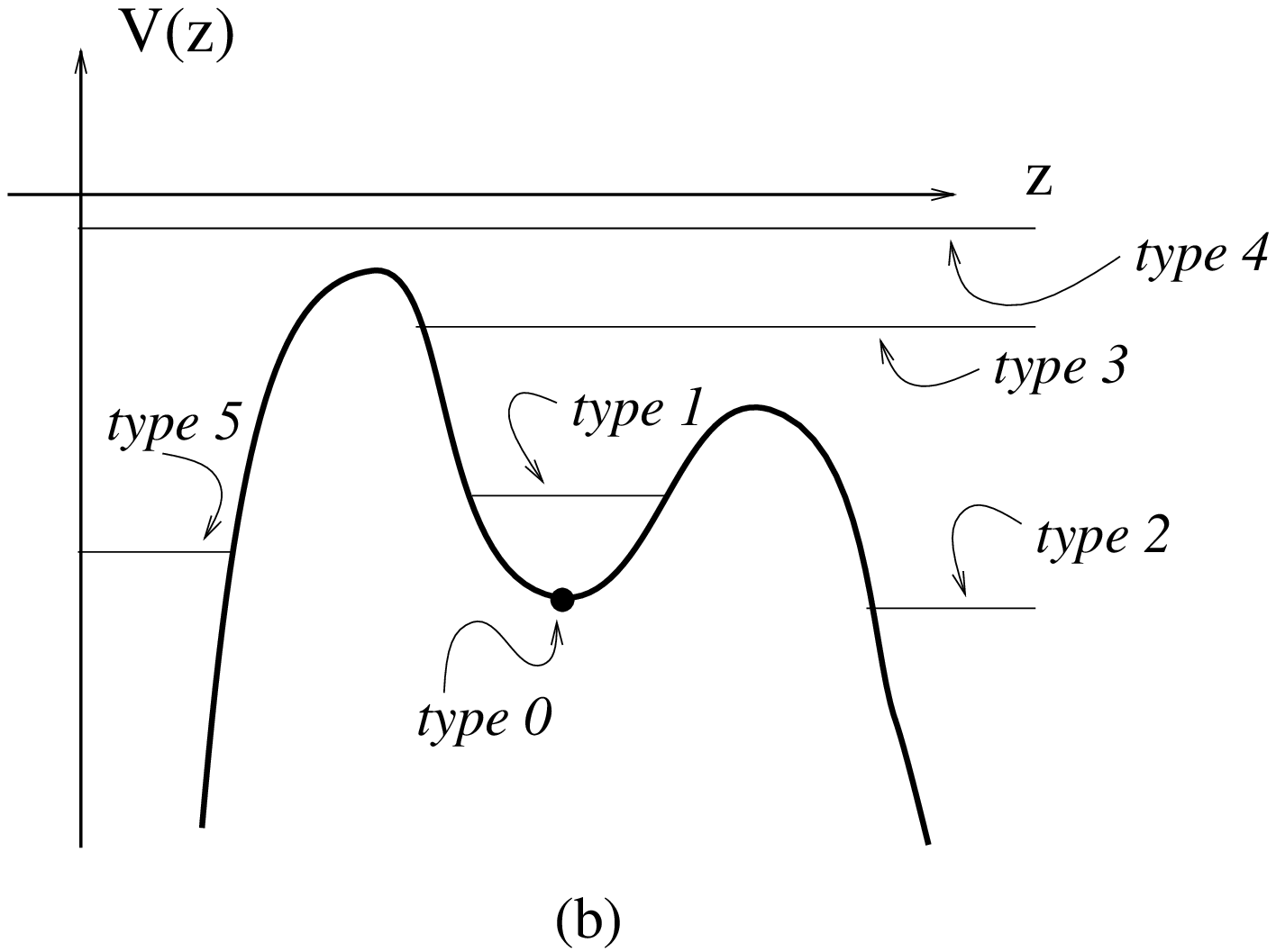}}\vskip -8pt
\noindent
For each of these levels there will be examples with $Q<M$, $Q=M$ and $Q>M$, 
depending on whether the
energy level lies above or below the RN horizon curve.  
In addition for $Q\le M$ whether, or not, a 
shell trajectory crosses the 
RN horizon line will lead to distinct causal structures.  
Since our primary interest is in shell trajectories that lead to regular black
holes, the important feature of the deSitter interior of the shell for us is its regularity.
However, we will also note for each class of trajectories whether the shell passes through a 
deSitter horizon and hence undergoes topological inflation in its interior.

\noindent
$\bullet$ {\it Type 0:} These are static shell configurations, 
generalizations of the static $\sigma=0$
solutions presented in the introduction for the timelike boundaries and shown in figures 
(1a-1c)\foot{Recall the matching in figure (1d) is along a spacelike shell.}.  
The Penrose diagrams are again those of figures (1a-c).  
Note that each shell covers only one singularity. For the maximally extended
regular black holes shown in the figures, we need a separate shell covering each of the 
singularities. Also note that from figure (17) we can confirm that the shell radius is
always smaller than the deSitter horizon radius for the static configurations.  Hence the
shell interior does not include an inflating region.

\vskip 0.1in\noindent
$\bullet$ {\it Type 1:} In type I trajectories
the shell oscillates between minimum and maximum radii. For $Q<M$, $Q=M$ and $Q>M$
there are three configurations, trajectories (d), (f) and (h) in table (1), 
that have Penrose diagrams resembling those for the static 
({\it i.e.} type 0) solutions
but with an infinite number of oscillations in the shell trajectory as it goes from past to future 
timelike infinity.  The existence of these trajectories demonstrates the stability of the static 
monopole shells to small oscillations.
In addition, for $Q<M$ and $Q=M$, there are oscillatory trajectories shown in
figure (4) that repeatedly pass out through white hole horizons 
and in through black hole horizons
as they traverse the fully extended spacetime. These are trajectories (e) and (g) in table (1).
\ifig\fone{Type $1$ trajectories. Here and below, the solid dot denotes the Reissner-Norstrom side of the 
shell. The opposite side of the spacetime diagram is replaced with the deSitter interior of the
shell.
The $Q=M$ example shows collapse of a single shell to form a regular black hole.
The $Q<M$ case would need additional shells to cover up all singularities in future of region I.}
{\epsfysize=1.75in \epsfbox{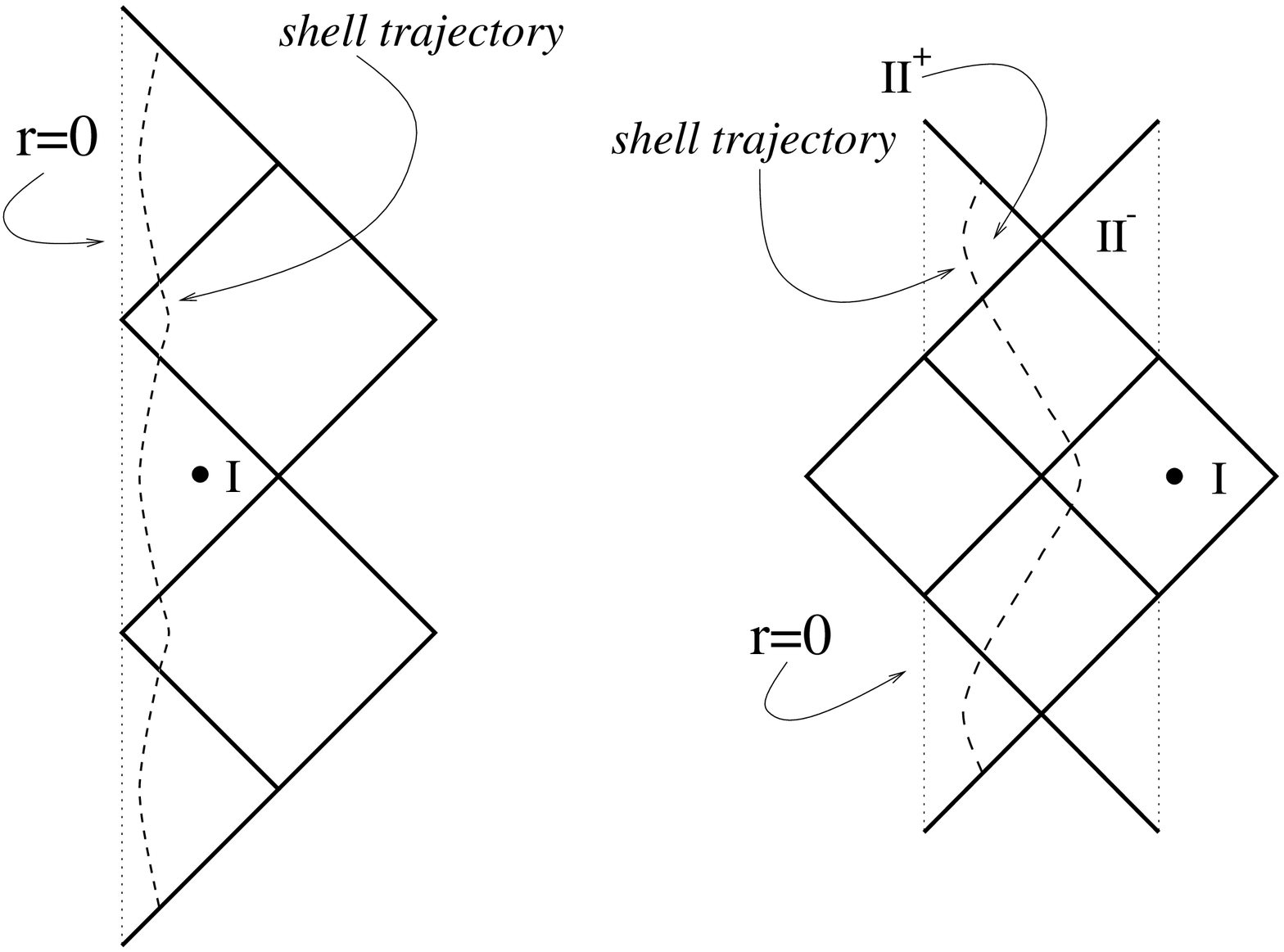}}\vskip -8pt
\noindent
These shells pass through region I at some time.  If we pick a spacelike surface in region I as an 
initial data surface, then an observer in region I sees a monopole collapsing to form a black hole.
The $Q=M$ case illustrates collapse of a single shell to form a regular black hole.
Note that in order to 
figure out which of the singularities in the $Q=M$ diagram 
is covered up by the shell, we use the fact that 
the normal $n_{RN}^r$ is positive everywhere along the path.  In region $II^+$ the normal is 
positive, {\it i.e.} moving into the Reissner-Nordstrom region is moving to larger values of the 
radial coordinate, while in region $II^{-}$ it is negative.
Finally, like the static configurations the shell radius is always smaller than the deSitter 
horizon radius for the type 1 trajectories.  Hence our regular black hole solutions do not include
inflating cores.

\vskip 0.1in\noindent
$\bullet$ {\it Type 2:} 
Trajectories of types $2$, $3$ and $4$ all start from $R=\infty$.  Note that
equation \firstorder\ implies that for large shell radii the sign of the normal $n^r_{RN}$ is 
always negative.  The normal vector by definition points away from the deSitter region into the 
Reissner-Nordstrom region.  The sign of the radial component being 
negative implies that moving from the
shell into the RN part of the spacetime, the radial coordinate is decreasing\foot{Although this 
appears contrary to equation, in fact the results continue to hold.}.  
If we have 
$n^r_{RN}<0$ in an asymptotically flat part of the spacetime, this indicates that the shell is on 
the left hand side of the Penrose diagram.

For type 2 trajectories, there is one type of shell path for $Q>M$ and two types 
each for $Q=M$ and $Q<M$.  
One set of possibilities for $Q>M$, $Q=M$ and $Q<M$ respectively is 
given by trajectories (i), (j) and (m) in table (1), which are shown in in figure (10).  
These trajectories have no
horizon crossings and always remain in a region with $n^r_{RN}<0$.  
The RN portions of the spacetime
diagrams are shown in figure (5). We see from figure (5) 
that the $Q\ge M$ spacetimes have no asymptotically flat region. The $Q<M$ spacetimes
contains the asymptotically flat region on the left hand side of the diagram. However, the monopole
shell never passes through this region and these spacetimes are still singular.
From the point of view of the deSitter interior, we can see from {\it e.g.} figure (17) that
trajectories of type 2 always start outside the deSitter horizon, pass in through the 
horizon and then cross out again.  
\ifig\fone{Type $2$ trajectories (i), (j) and (m) respectively. 
In each case the RN region is to the left of the shell trajectory since $n_{RN}^r<0$.}
{\epsfysize=1.4in \epsfbox{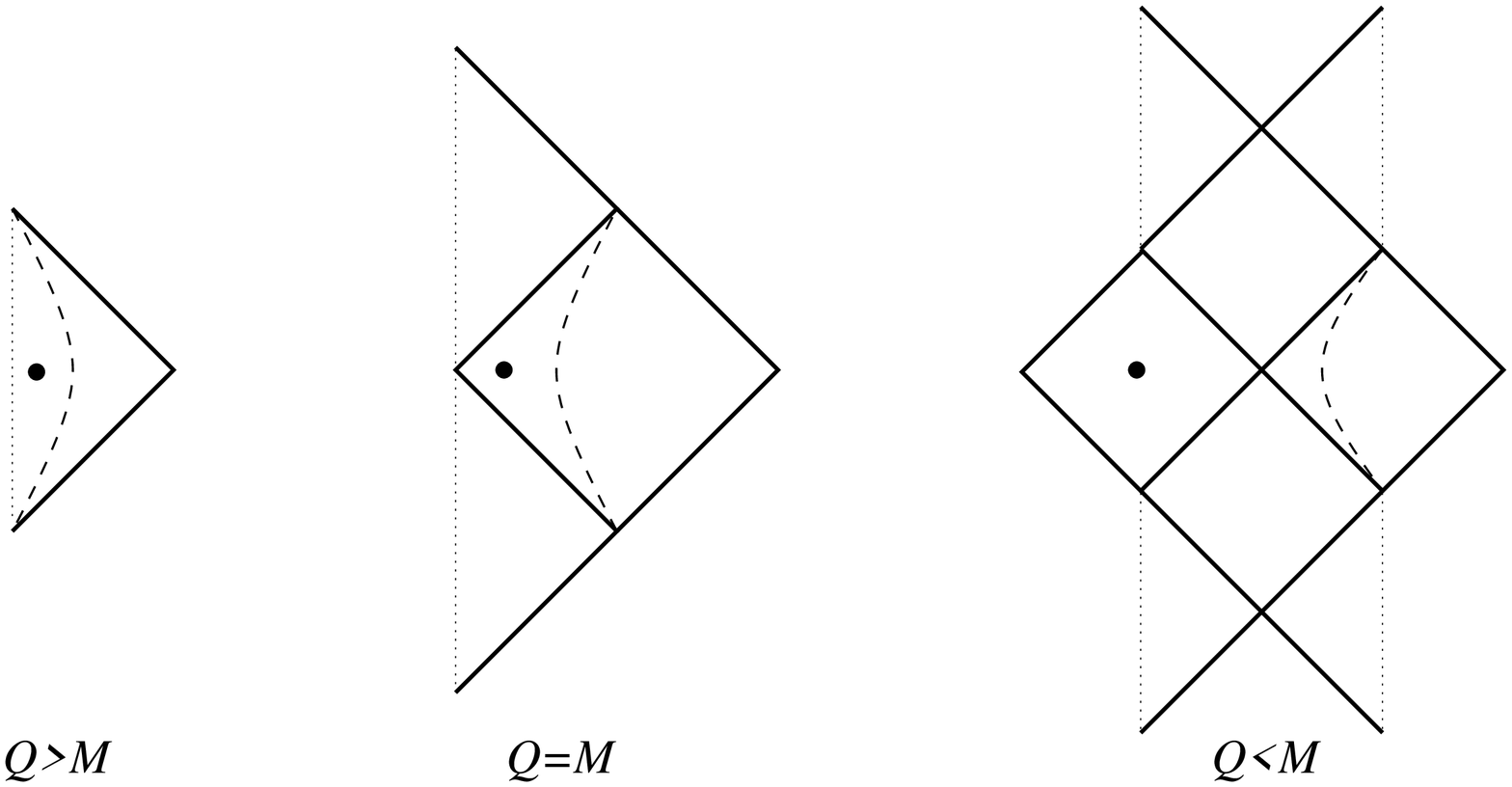}}
%

The second set of possibilities for $Q=M$ and $Q<M$, trajectories (k) and (n) in table (1),  
cross horizons but still have $n^r_{RN}<0$ everywhere along the path.  
These are shown in figure (6).  Again we see that the $Q=M$ spacetimes have no asymptotically flat 
region and that in the $Q<M$ spacetimes the monopole shell is never passes through the
asymptotically flat region.
\ifig\fone{Type $2$ trajectories (k) and (n). In each case the RN region 
is to the left of the shell trajectory.}
{\epsfysize=1.4in \epsfbox{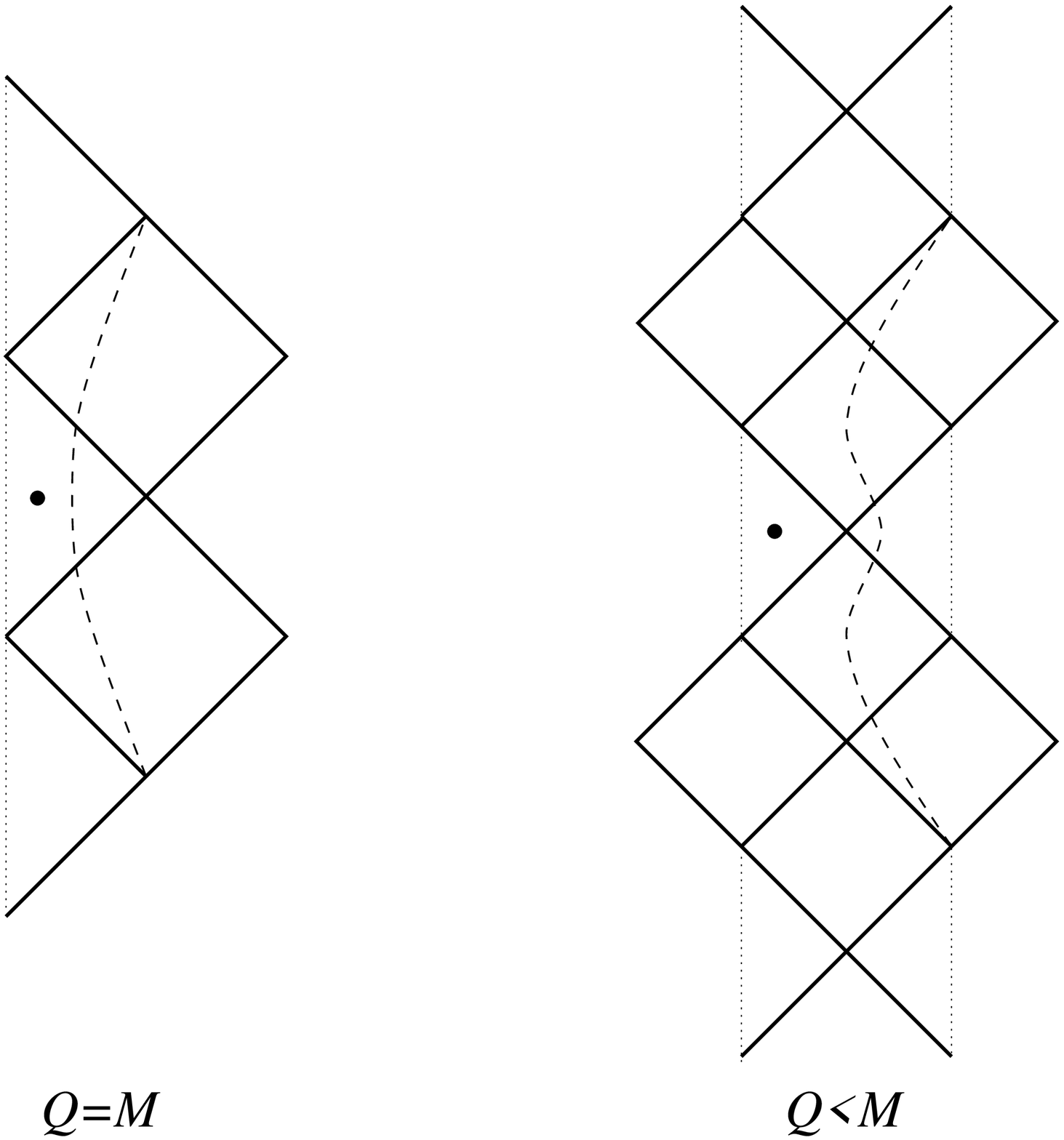}}\vskip -8pt
%

\vskip 0.1in\noindent
$\bullet$ {\it Type 3:} There are one $Q=M$ trajectory and two $Q<M$ trajectories, 
(l), (o) and (p) in table (1) respectively. 
Trajectories (l) and (o) have the form shown in figure 
(6) above, while trajectory (p) has the form shown in figure (7).  The qualitative
differences between trajectories (o) and (p) involve only how the sign of the normal 
changes along the curve.  Again in the $Q<M$ spacetimes the shell is never passes through an
asymptotically flat region of the spacetime.
The type 3 trajectories viewed from the deSitter interior also always pass in and then out of
the deSitter horizon.
\ifig\fone{Type 3 trajectory (p).  Again the RN region is to the left of the curve.}
{\epsfysize=1.5in \epsfbox{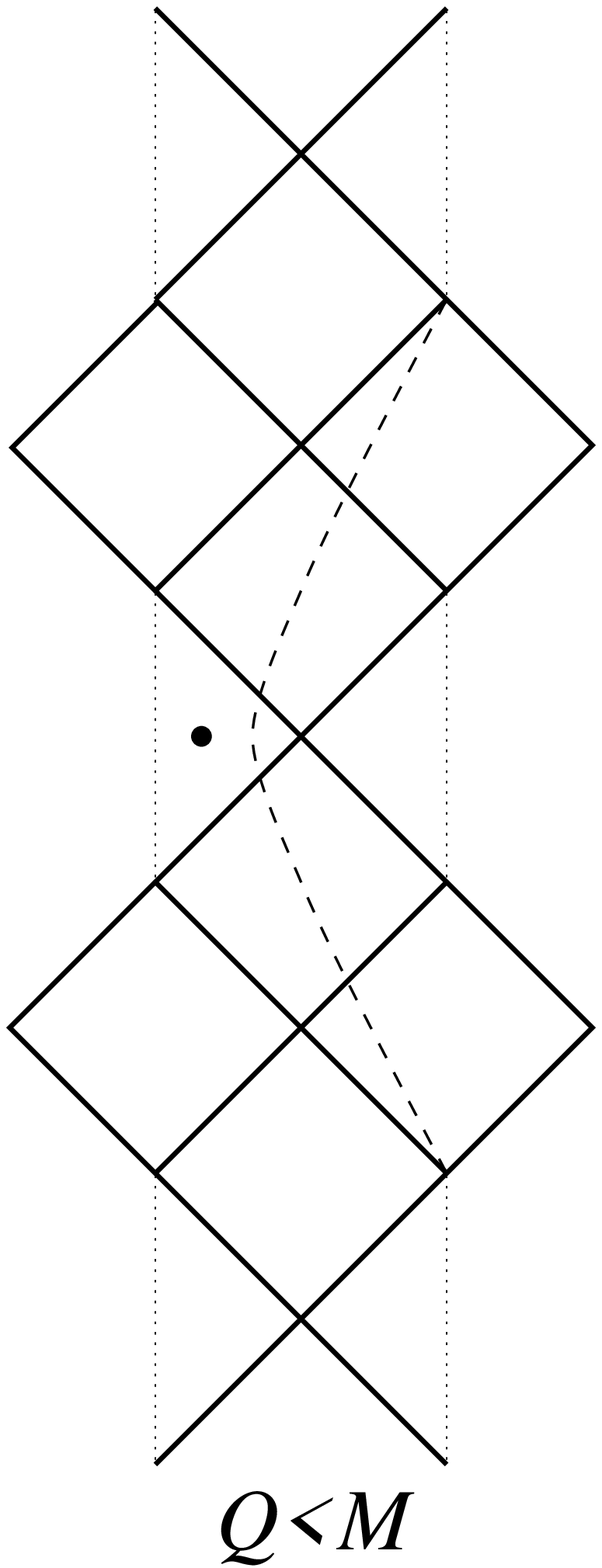}}\vskip -8pt
%

\vskip 0.1in\noindent
$\bullet$ {\it Type 4:} 
These trajectories move in from infinity and hit the curvature singularity at 
$R=0$ in finite proper time.  
There is one possibility for $Q>M$ - trajectory (q);
one for $Q=M$ - trajectory (r); and two for $Q<M$ - trajectories (s) and (t),
all shown in figure (8).
The Reissner-Nordstrom portions of these spacetimes are all singular and again 
are uninteresting for our purposes. Again the $Q\ge M$ spacetimes are not asymptotically
flat and the monopole shell does not pass through the asymptotically
flat region of the $Q<M$ spacetimes.
In these spacetimes the shell passes in through the deSitter horizon, but then 
never exits. Thus they include a past deSitter horizon, but not a future horizon.
\ifig\fone{Type 4 trajectories (q), (r), (s) and (t).}
{\epsfysize=1.5in \epsfbox{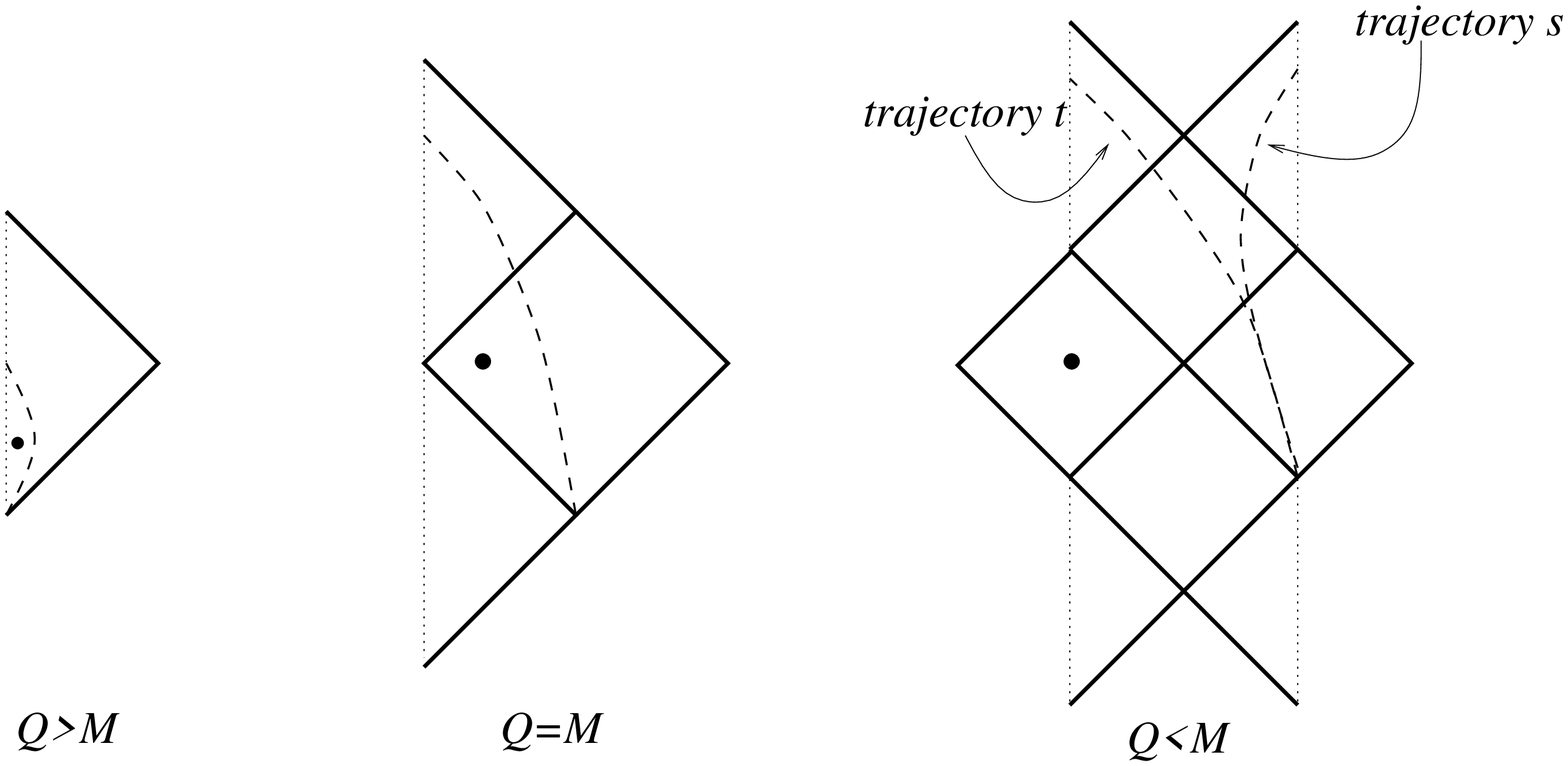}}\vskip -8pt
\vskip 0.1in\noindent
$\bullet$ {\it Type 5:} Finally, type 5 trajectories (a), (b) and (c) 
leave $R=0$ and return within finite proper time.
All of these spacetimes contain naked singularities.
The Reissner-Nordstrom sides of the spacetime diagrams are shown in figure (9).
\ifig\fone{Type 5 trajectories (a), (b) and (c)
depart and return to the curvature singularity 
$R=0$ in finite proper time.}
{\epsfysize=1.5in \epsfbox{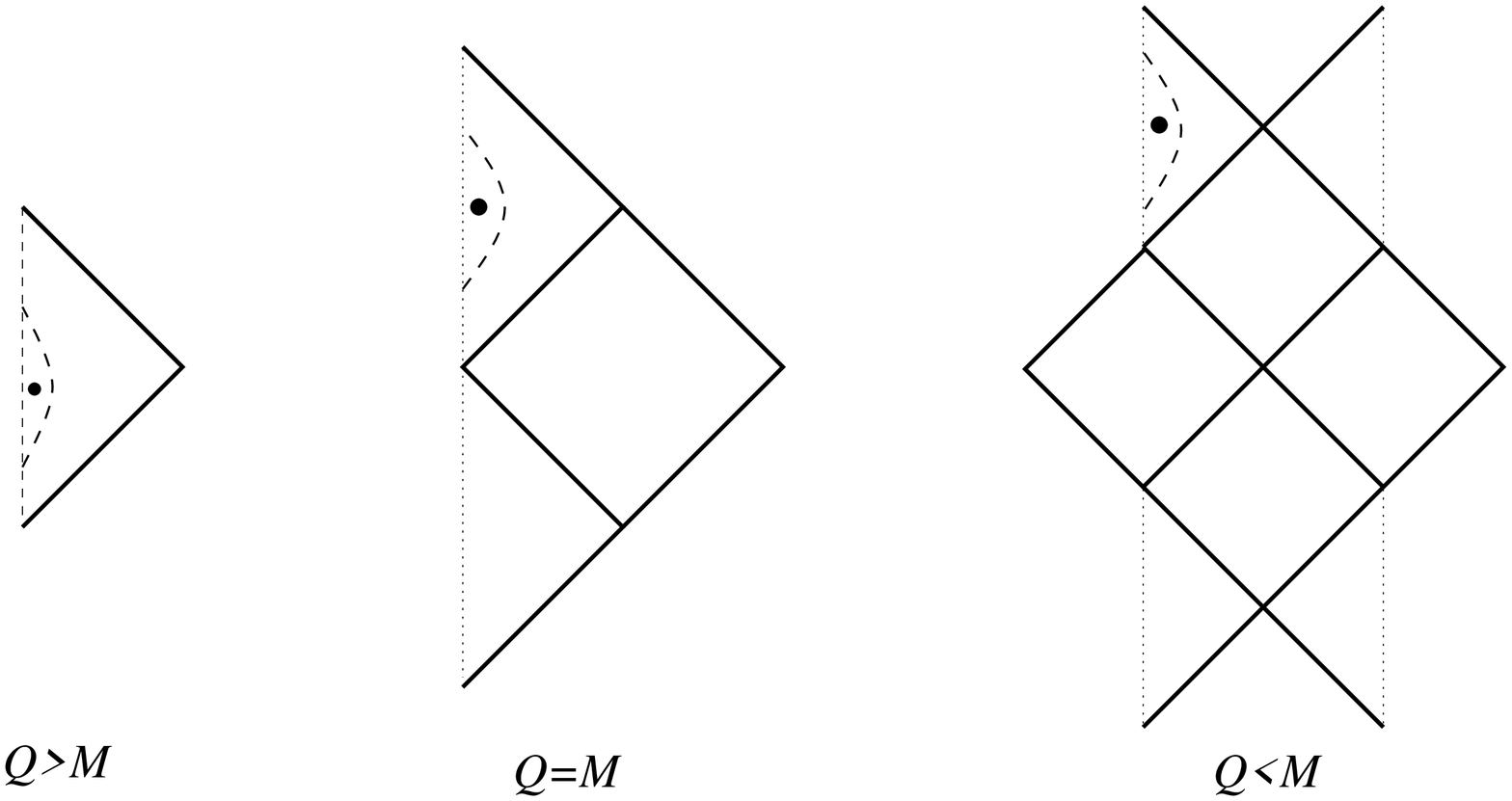}}\vskip -8pt

\vfill\supereject

\appendix{A}{}
Table (1) summarizes the qualitatively different possible shell trajectories when viewed from 
the Reissner-Nordstrom part of the spacetime.  The various paths (st1, st2, st3) and (a-t) are labeled
on the potential diagrams in figures (9-16).
\vskip 0.15in

\centerline{\vbox{\offinterlineskip
\hrule
\halign{&\vrule#&
  \strut\quad\hfill#\quad\cr
height2pt&\omit&&\omit&&\omit&&\omit&&\omit&\cr
&path\hfill&&range of z\hfill&&Sign of $\nout$\hfill&& Q and M\hfill&&Horizon
Crossing&\cr
height2pt&\omit&&\omit&&\omit&&\omit&&\omit&\cr
\noalign{\hrule}
height2pt&\omit&&\omit&&\omit&&\omit&&\omit&\cr
&a&&$0 < z \le \zmax$&&$-$&&$Q > M$&&*&\cr
&b&&$0 < z \le \zmax$&&$-$&&$Q = M$&&no&\cr
&c&&$0 < z \le \zmax$&&$-$&&$Q < M$&&no&\cr
height2pt&\omit&&\omit&&\omit&&\omit&&\omit&\cr
\noalign{\hrule}
height2pt&\omit&&\omit&&\omit&&\omit&&\omit&\cr
&st1&&$z = \zst$&&$+$&&$Q > M$&&*&\cr
&st2&&$z = \zst$&&$+$&&$Q = M$&&no&\cr
&st3&&$z = \zst$&&$+$&&$Q < M$&&no&\cr
height2pt&\omit&&\omit&&\omit&&\omit&&\omit&\cr
\noalign{\hrule}
height2pt&\omit&&\omit&&\omit&&\omit&&\omit&\cr
&d&&$\zmin \le z \le \zmax$&&$+$&&$ Q > M$&&*&\cr
&e&&$\zmin \le z \le \zmax$&&$+$&&$ Q = M$&&yes&\cr
&f&&$\zmin \le z \le \zmax$&&$+$&&$ Q = M$&&no&\cr
&g&&$\zmin \le z \le \zmax$&&$+$&&$ Q < M$&&yes&\cr
&h&&$\zmin \le z \le \zmax$&&$+$&&$ Q < M$&&no&\cr
height2pt&\omit&&\omit&&\omit&&\omit&&\omit&\cr
\noalign{\hrule}
height2pt&\omit&&\omit&&\omit&&\omit&&\omit&\cr
&i&&$\zmin \le z < \infty$&&$-$&&$ Q > M$&&*&\cr
&j&&$\zmin \le z < \infty$&&$-$&&$Q = M$&&**&\cr
&k&&$\zmin \le z < \infty$&&$-$&&$Q = M$&&yes&\cr
&l&&$\zmin \le z < \infty$&&$+ -$&&$Q = M$&&yes&\cr
&m&&$\zmin \le z < \infty$&&$-$&&$Q < M$&&**&\cr
&n&&$\zmin \le z < \infty$&&$-$&&$Q < M$&&yes&\cr
&o&&$\zmin \le z < \infty$&&$+ -$&&$Q < M$&&yes&\cr
&p&&$\zmin \le z < \infty$&&$- + -$&&$Q < M$&&yes&\cr
height2pt&\omit&&\omit&&\omit&&\omit&&\omit&\cr
\noalign{\hrule}
height2pt&\omit&&\omit&&\omit&&\omit&&\omit&\cr
&q&&$0 < z < \infty$&&$- $&&$Q > M$&&*&\cr
&r&&$0 < z < \infty$&&$-$&&$Q = M$&&yes&\cr
&s&&$0 < z < \infty$&&$- + -$&&$ Q < M$&&yes&\cr
&t&&$0 < z < \infty$&&$-$&&$Q < M$&&yes&\cr
height2pt&\omit&&\omit&&\omit&&\omit&&\omit&\cr}
\hrule}}
\centerline{\it Table 1. Character of each path of the shell-
Reissner-Nordstr\"om part}
\midinsert\narrower\narrower{
{\ninerm * $Q > M$ - no black hole horizons.\hfill\break
** There is no horizon present in the spacetime, since they (it) occur(s) inside the
   shell.}
}
\endinsert\vskip5mm
\vfill\eject

Table (2) summarizes the qualitatively different possible shell trajectories when viewed from 
the deSitter part of the spacetime.  The various paths (ST) and (A-H) are labeled
on the potential diagrams in figures (17-21).
\vskip 0.15in

\centerline{\vbox{\offinterlineskip
\hrule
\halign{&\vrule#&
  \strut\quad\hfill#\quad\cr
height2pt&\omit&&\omit&&\omit&&\omit&\cr
&Path\hfill&&Range of z\hfill&&Sign of $\nin$\hfill&&Horizon Crossing&\cr 
height2pt&\omit&&\omit&&\omit&&\omit&\cr
\noalign{\hrule}
height2pt&\omit&&\omit&&\omit&&\omit&\cr
&A&&$0 < z \le \zmax$&&$ - $&&*&\cr
height2pt&\omit&&\omit&&\omit&&\omit&\cr
\noalign{\hrule}
height2pt&\omit&&\omit&&\omit&&\omit&\cr
&ST&&$z=\zst$&&$ + $&&*&\cr
height2pt&\omit&&\omit&&\omit&&\omit&\cr
\noalign{\hrule}
height2pt&\omit&&\omit&&\omit&&\omit&\cr
&B&&$\zmin \le z \le \zmax$&&$ + $&&*&\cr
height2pt&\omit&&\omit&&\omit&&\omit&\cr
\noalign{\hrule}
height2pt&\omit&&\omit&&\omit&&\omit&\cr
&C&&$\zmin \le z <\infty$&&$ + $&&yes&\cr
&D&&$\zmin \le  z <\infty$&&$ + - $&&yes&\cr
&E&&$\zmin \le z <\infty$&&$ - $&&yes&\cr
height2pt&\omit&&\omit&&\omit&&\omit&\cr
\noalign{\hrule}
height2pt&\omit&&\omit&&\omit&&\omit&\cr
&F&&$0 < z < \infty$&&$ - $&&yes&\cr
&G&&$0 < z < \infty$&&$ - + $&&yes&\cr
&H&&$0 < z < \infty$&&$ - + - $&&yes&\cr
height2pt&\omit&&\omit&&\omit&&\omit&\cr}
\hrule}}
\centerline{\it Table 2.\quad Character of each path of the shell -
de Sitter part}
\midinsert\narrower\narrower\narrower{
\noindent{\ninerm * The de Sitter horizon does not exist in these spacetimes since it lies 
outside the shell.}
}
\endinsert\vskip3mm

\vfill\eject

Table (3) lists the ranges of parameters $\alpha$ and $\gamma$ that are relevant to the 
configurations of potentials and horizon curves shown in figures (10)-(19).

\vskip 0.15in

\vbox{\offinterlineskip
\hrule
\halign{&\vrule#&
  \strut\quad\hfill#\quad\cr
height2pt&\omit&&\omit&&\omit&&\omit&\cr
&$\alpha$\hfill&&$\gamma ^2$\hfill&&$V-\vds$\hfill&&$V-\vrn$&\cr
height2pt&\omit&&\omit&&\omit&&\omit&\cr
\noalign{\hrule}
height2pt&\omit&&\omit&&\omit&&\omit&\cr
&$\alpha < \left({1 \over 2}\right)^{4/3}$&&$\gamma^2 < 2$&&\I&&1&\cr\
&\omit&&$\gamma^2 \ge 2$&&\II&&1&\cr
height2pt&\omit&&\omit&&\omit&&\omit&\cr
\noalign{\hrule}
height2pt&\omit&&\omit&&\omit&&\omit&\cr
&$\left({1 \over 2}\right)^{4/3} <\alpha <\alphao$&&$\gamma^2 < 2$&&\I&&2,3,4&
\cr
&\omit&&$2 \le \gamma^2 < \gammamax$&&\II&&2&\cr
&\omit&&$\gamma^2 \ge \gammamax$&&\III&&5&\cr
height2pt&\omit&&\omit&&\omit&&\omit&\cr
\noalign{\hrule}
height2pt&\omit&&\omit&&\omit&&\omit&\cr
&$\alpha = \alphao$&&$\gamma^2 < 2(=\gammamax)$&&\I&&2,3,4&\cr
&\omit&&$\gamma^2 \ge 2(=\gammamax)$&&\III&&5&\cr
height2pt&\omit&&\omit&&\omit&&\omit&\cr
\noalign{\hrule}
height2pt&\omit&&\omit&&\omit&&\omit&\cr
&$\alphao < \alpha < {3 \over {4^{4/3}}}$&&$\gamma^2 < \gammamax$&&\I&&2,3,4&\cr
&\omit&&$\gammamax \le \gamma^2 < 2$&&\IV&&5&\cr
&\omit&&$\gamma^2 > 2$&&\III&&5&\cr
height2pt&\omit&&\omit&&\omit&&\omit&\cr
\noalign{\hrule}
height2pt&\omit&&\omit&&\omit&&\omit&\cr
&$\alpha > {3 \over {4^{4/3}}}$&&$\gamma^2 < 2 \left(1-{27 \over 
{256 \alpha^3}}
\right)$&&V&&6,7,8&\cr
&\omit&&$2\left(1-{27 \over{256 \alpha^3}}\right) < \gamma^2 < 2$&&\IV&&6,7,8&
\cr
&\omit&&$\gamma^2 \ge 2$&&\III&&6,7,8&\cr
height2pt&\omit&&\omit&&\omit&&\omit&\cr}
\hrule}
\centerline{\it Table 3. }
\medskip
*For $\alpha = \alphao (=0.40418572)$, $\gammamax =2$.

\vfill\eject


%
\ifig\fone{Reissner-Nordstrom configuration 1.}
{\epsfysize=1.45in \epsfbox{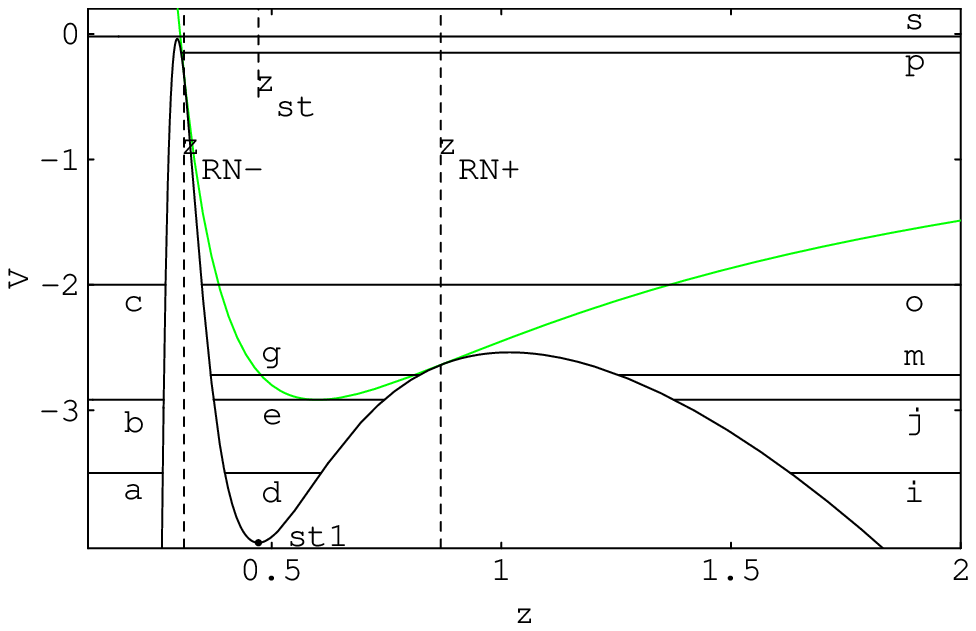}}
\vskip -8pt
\ifig\fone{Reissner-Nordstrom configuration 2.}
{\epsfysize=1.45in \epsfbox{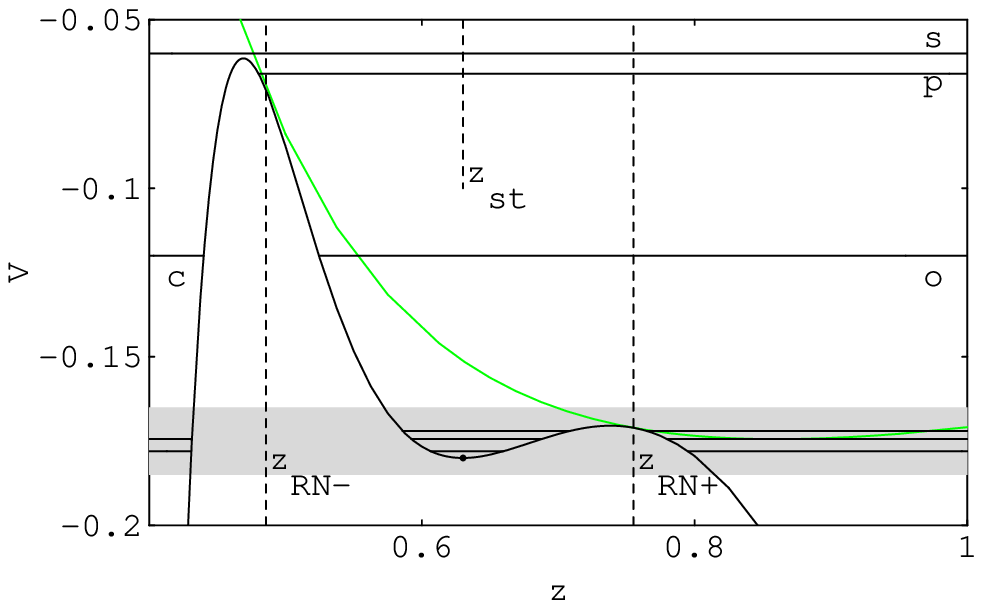}\epsfysize=1.45in \epsfbox{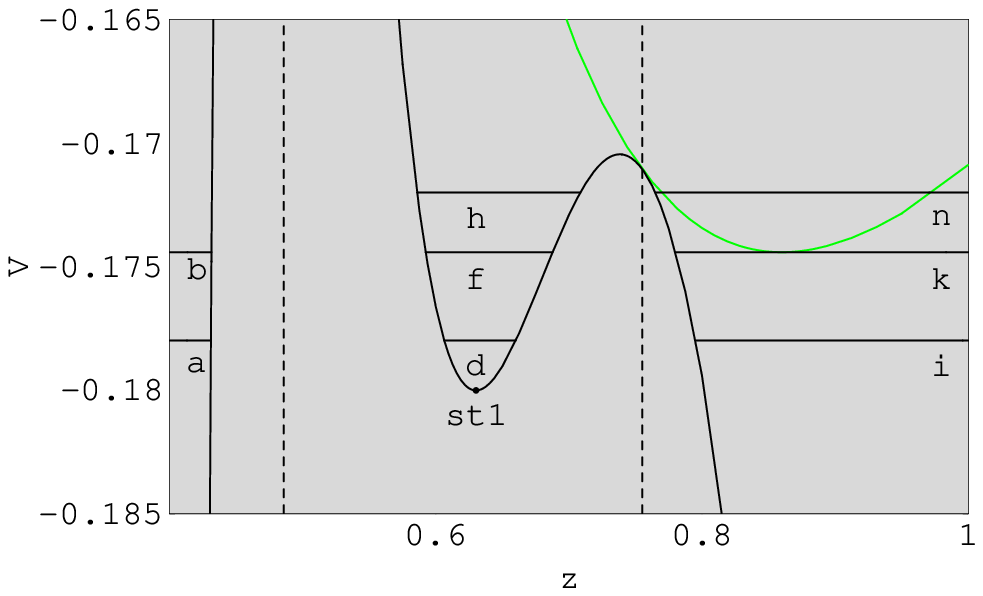}}
\vskip -8pt
\ifig\fone{Reissner-Nordstrom configuration 3.}
{\epsfysize=1.45in \epsfbox{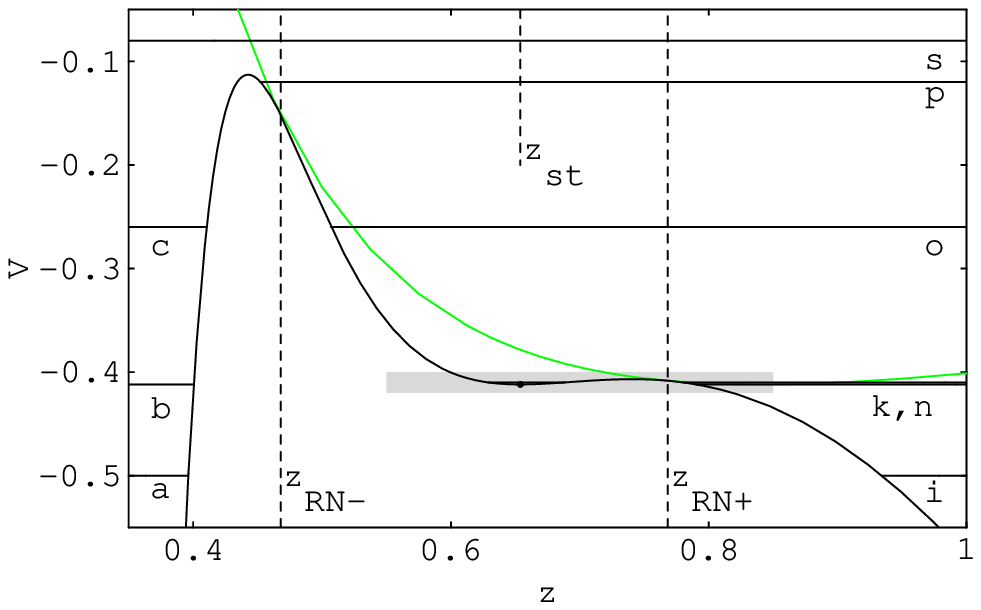}\epsfysize=1.45in \epsfbox{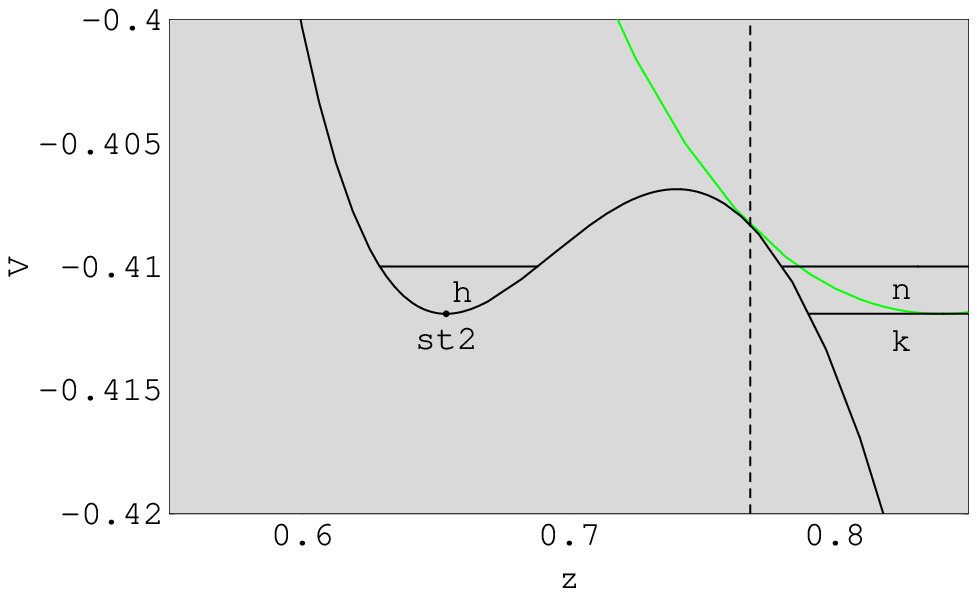}}
\vskip -8pt
\ifig\fone{Reissner-Nordstrom configuration 4.}
{\epsfysize=1.45in \epsfbox{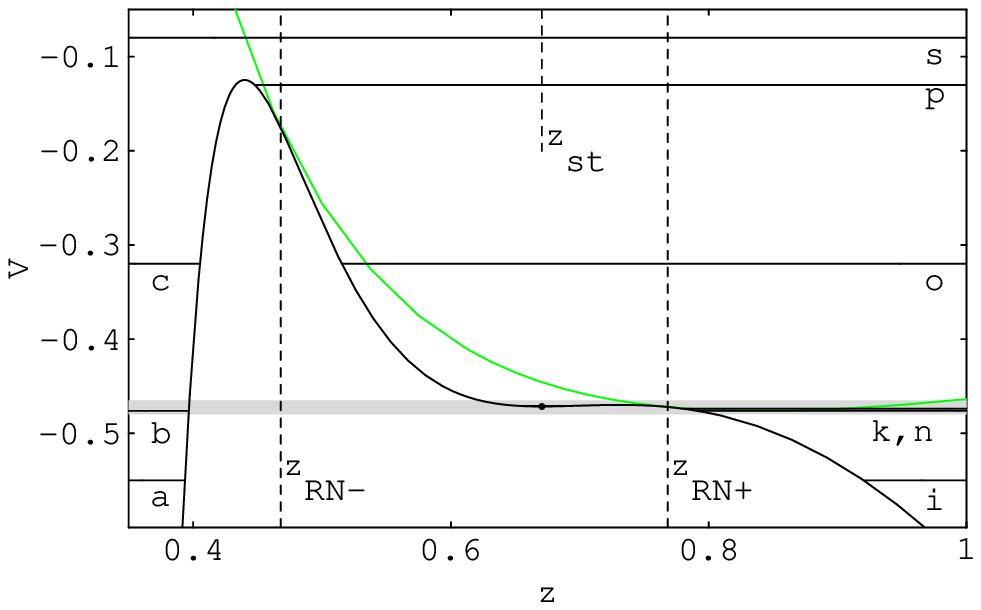}\epsfysize=1.45in \epsfbox{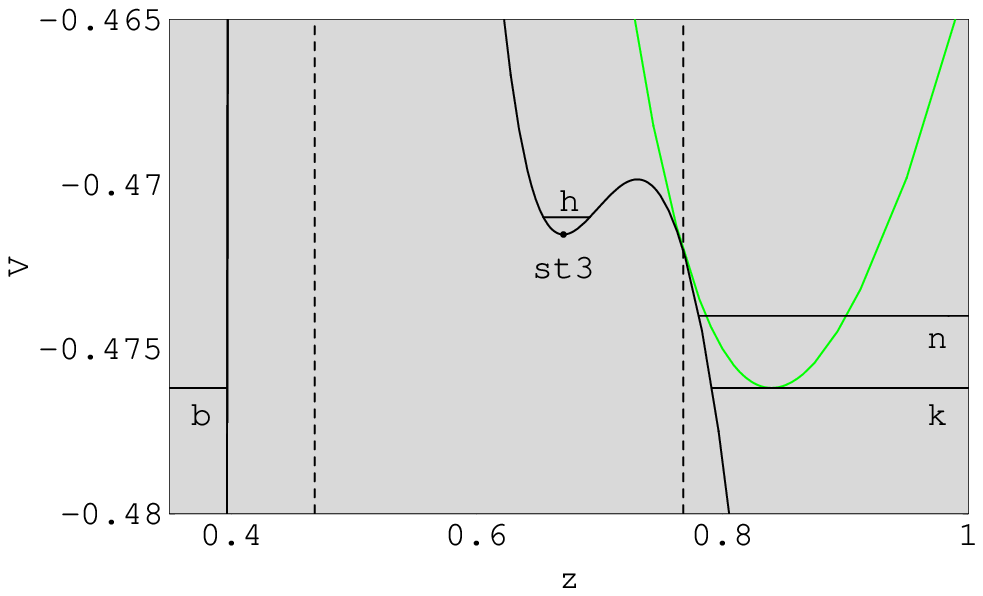}}
\vskip -8pt
\ifig\fone{Reissner-Nordstrom configuration 5.}
{\epsfysize=1.45in \epsfbox{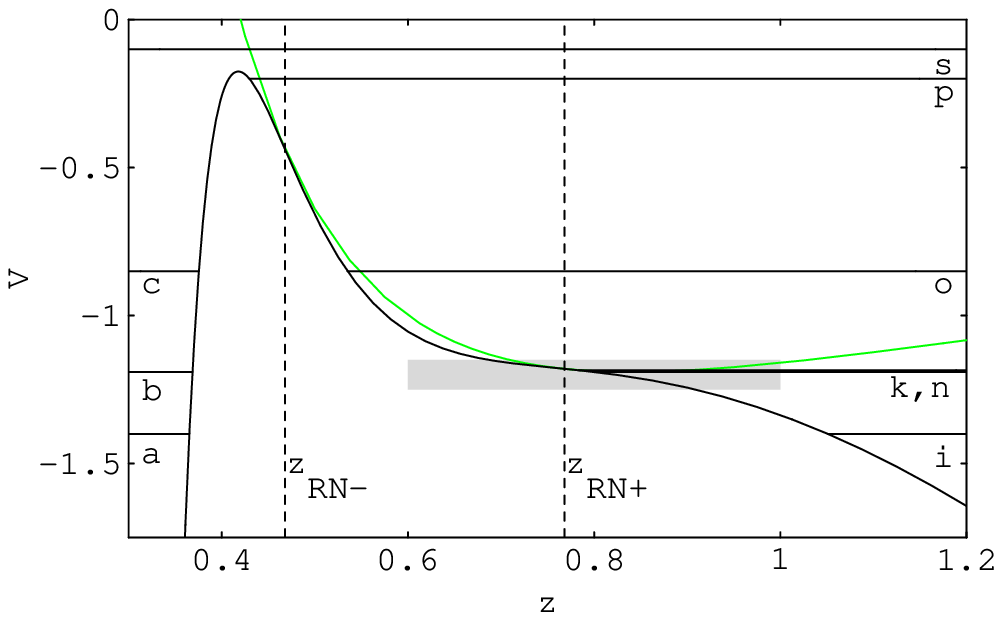}\epsfysize=1.45in \epsfbox{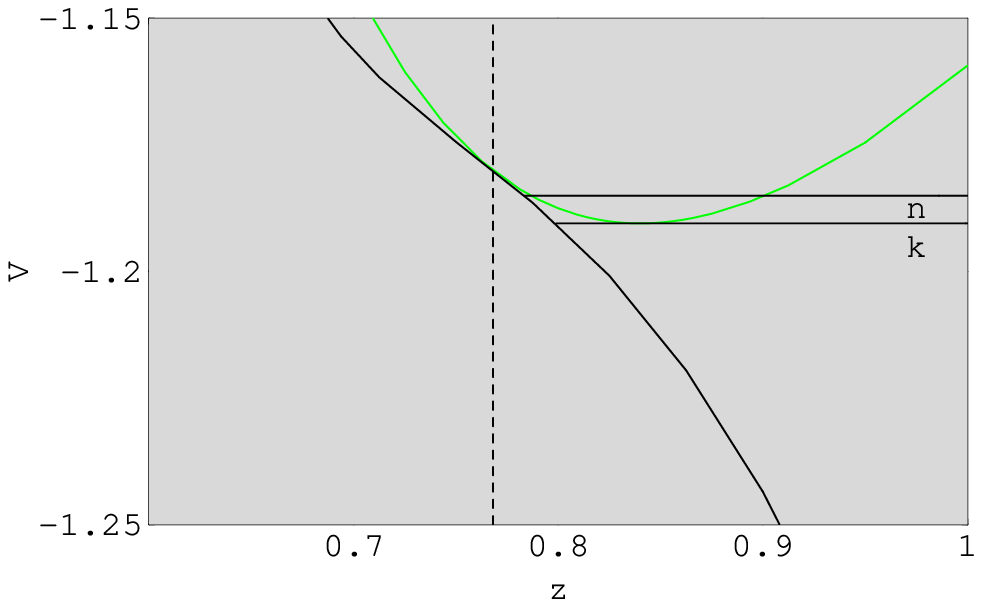}}
\vskip -8pt
\ifig\fone{Reissner-Nordstrom configuration 6 and 7.}
{\epsfysize=1.45in \epsfbox{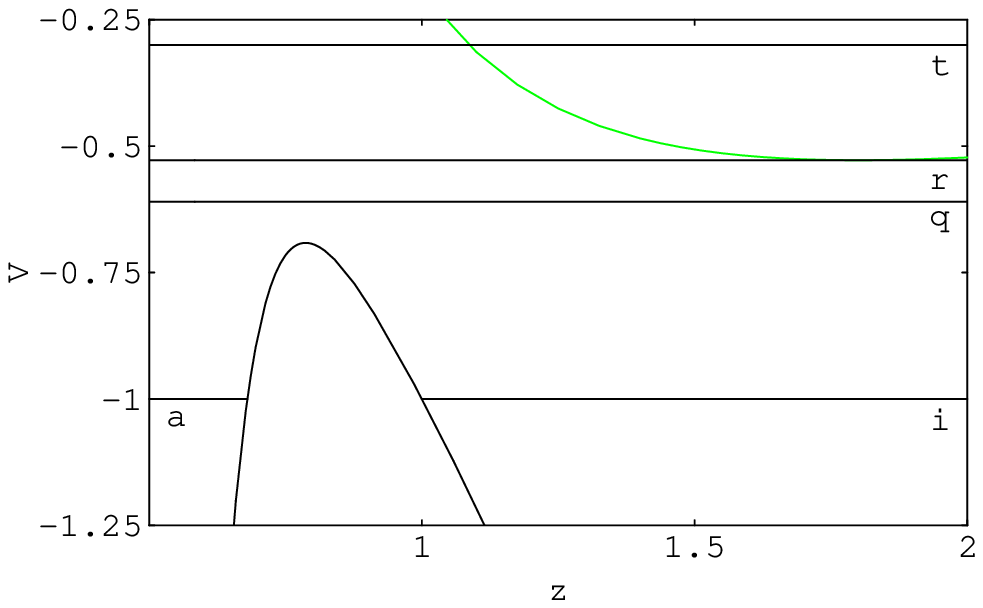}\epsfysize=1.45in \epsfbox{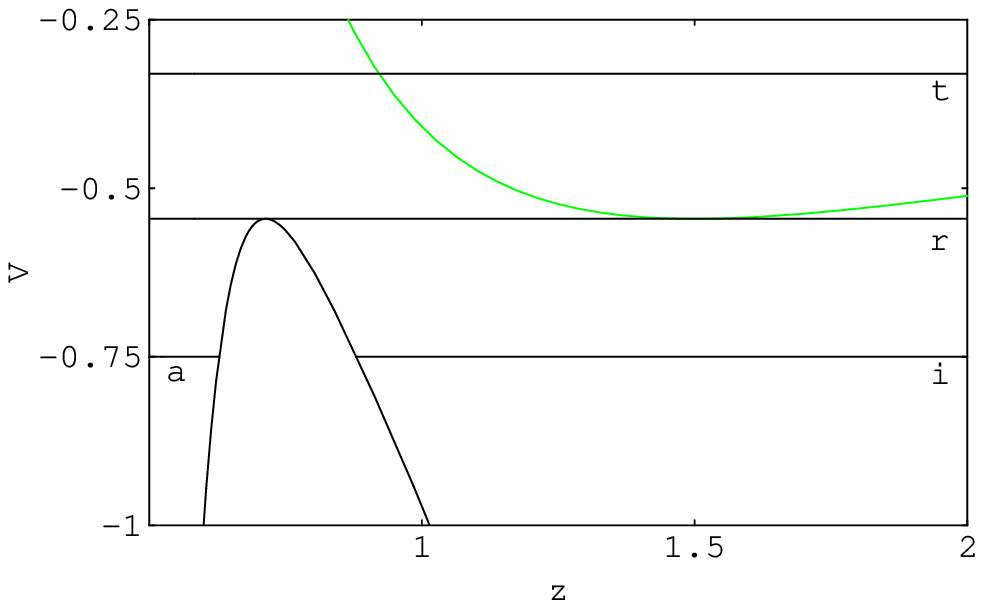}}
\vskip -8pt
\ifig\fone{Reissner-Nordstrom configuration 8.}
{\epsfysize=1.45in \epsfbox{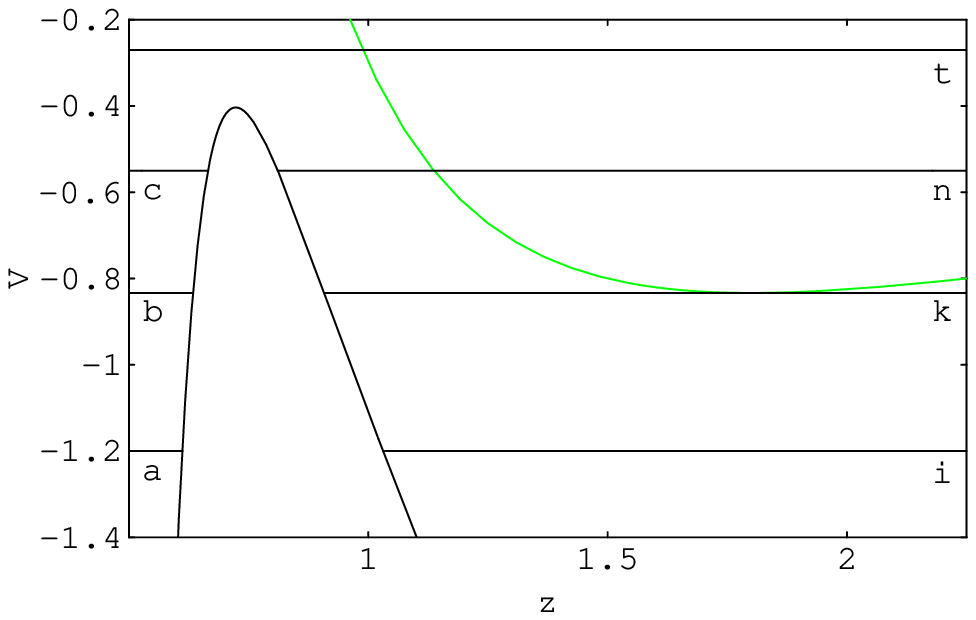}}
\vskip -8pt
%


%
\ifig\fone{deSitter configurations I and II.}
{\epsfysize=1.45in \epsfbox{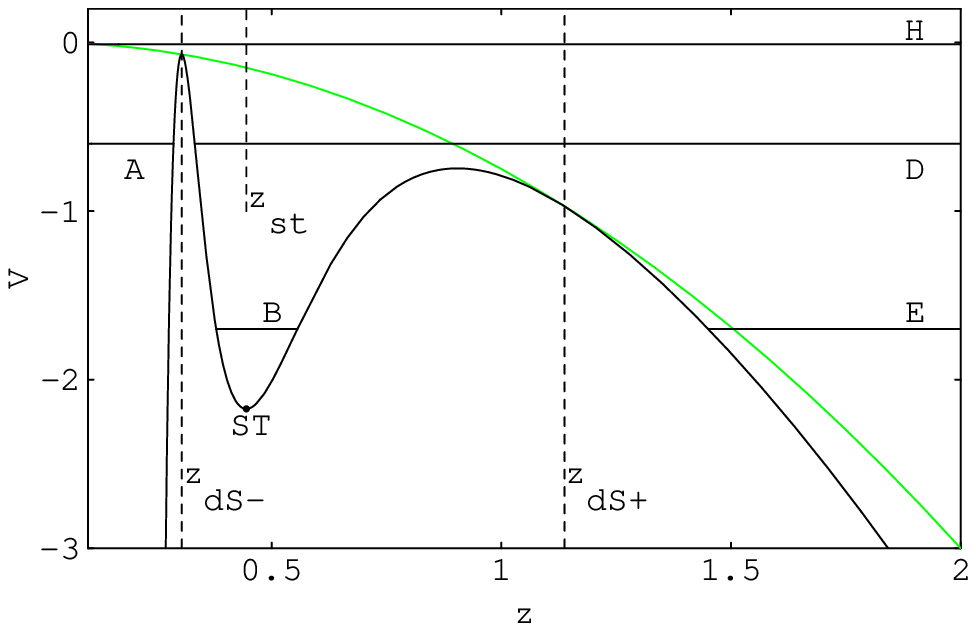}\epsfysize=1.45in \epsfbox{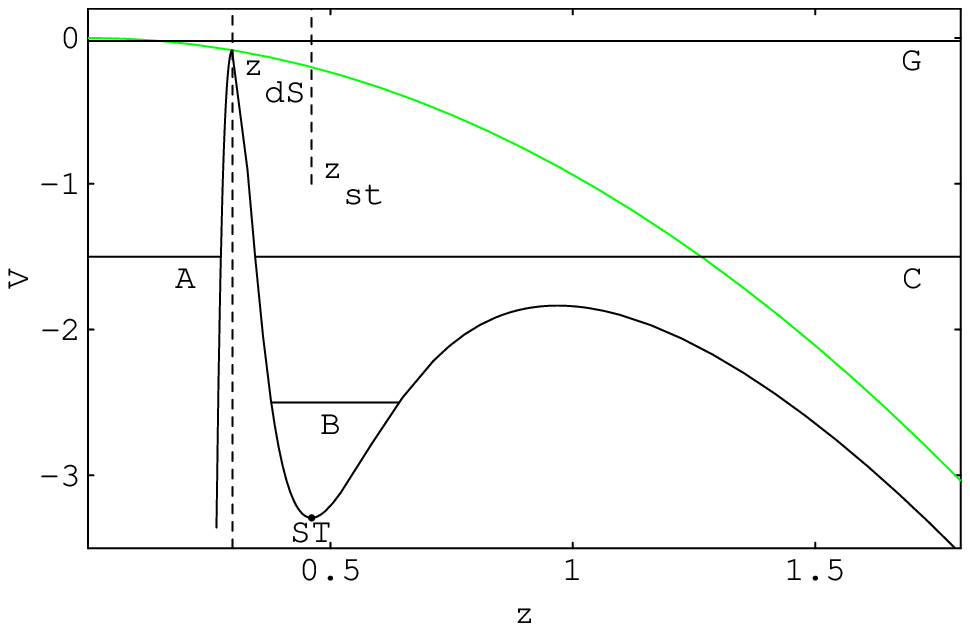}}
\vskip -8pt
\ifig\fone{deSitter configurations III and IV.}
{\epsfysize=1.45in \epsfbox{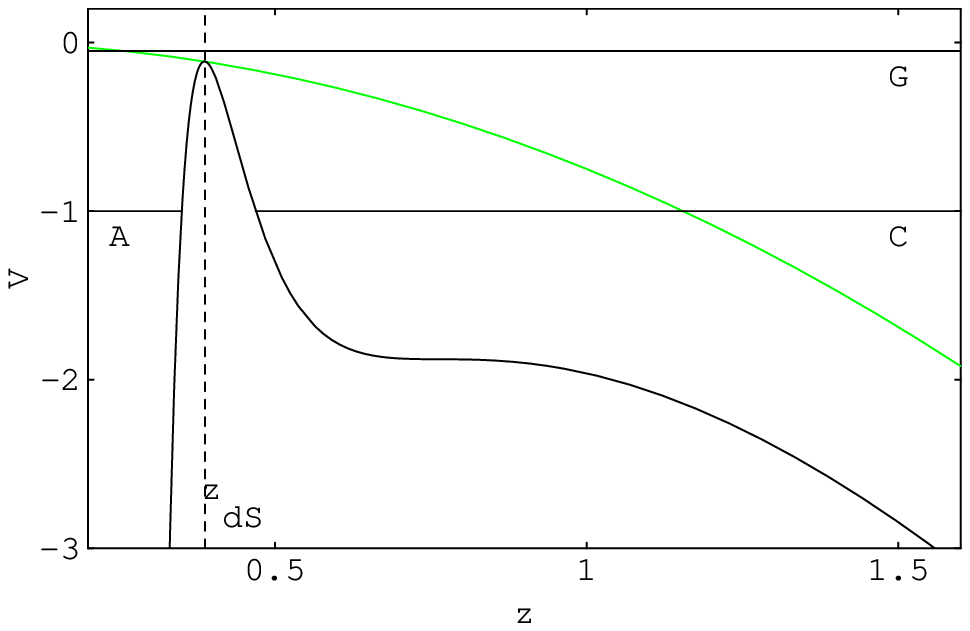}\epsfysize=1.45in \epsfbox{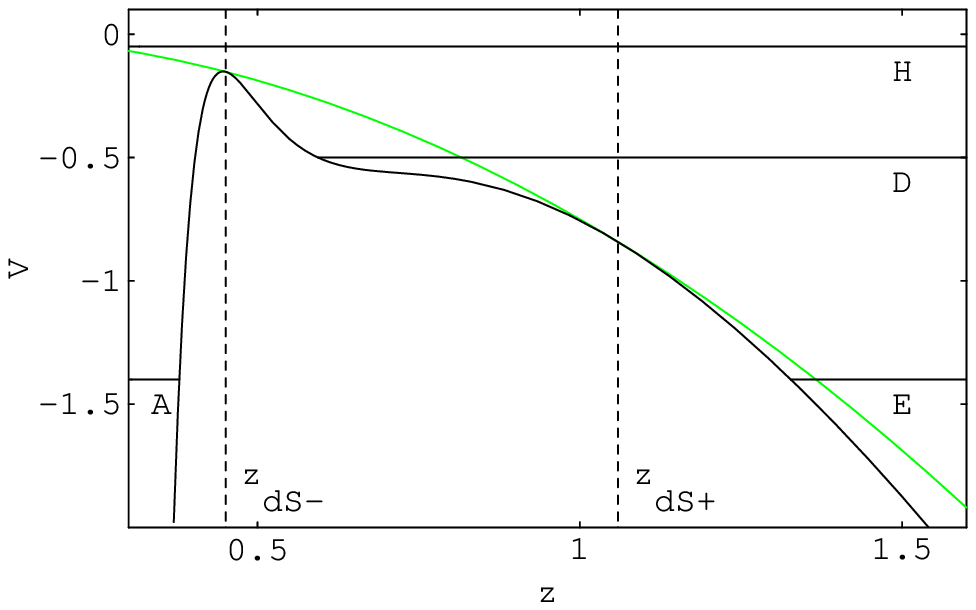}}
\vskip -8pt
\ifig\fone{deSitter configuration V.}
{\epsfysize=1.45in \epsfbox{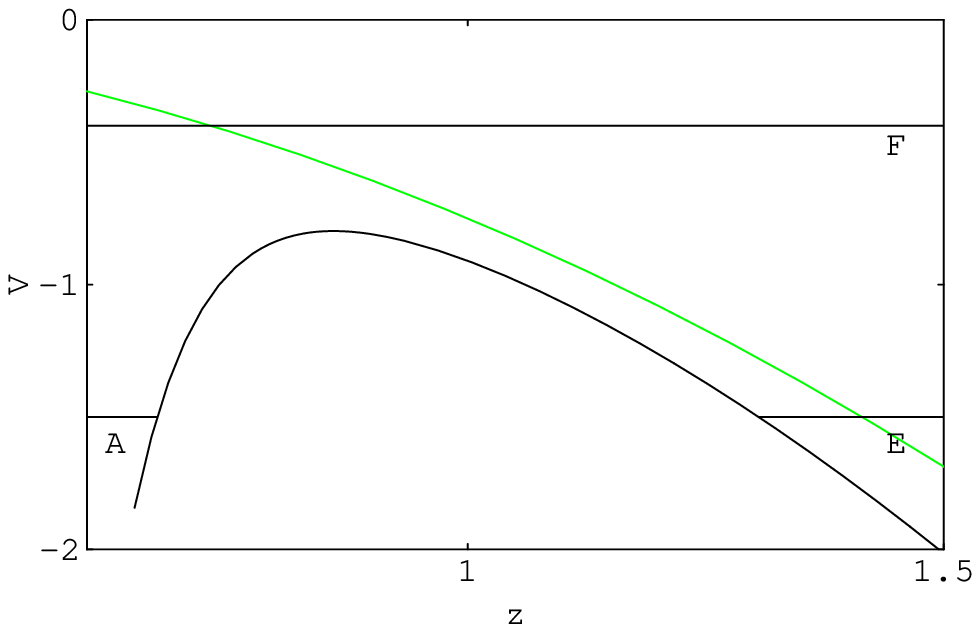}}
\vskip -8pt

\listrefs\end

\newsec{Possible Solutions}
The character of each path is determined by the potential shape and the 
position of the horizons set by $\alpha$, $\gamma^2$ and the energy $E$. 
It is shown in the table. Now the following is the brief explanation on each 
path with spacetime diagrams.

{\bf\expandafter{\sl\subsec{Bounded Solution}}}
 The shell radius begins at zero and 
increases to the maximum value. Inside the shell, we have de Sitter spacetime 
and the radial normal is negative for the entire path. Outside the shell, 
we have three types of spacetime according to the relationship between $Q$ and 
$M$. The sign of the radial component of normal is negative in all the cases.

\eject

\midinsert{
\centerline{\epsfbox{bounded.eps}}
{\centerline{ De Sitter region\qquad\qquad\  Reissner-Nordstr\"om Region}}
}
\endinsert

??? $Q^2 > M^2$  case, the normal is negative. Nonsense???

{\bf\expandafter\sl\subsec{ Static Solution}} 
 When the potential has the local minimum, 
we have a stable static solution. As the bounded solution case, we have one 
kind of inside region; de Sitter spacetime with no horizon-line crossing and 
positive radial component of normal. Outside the shell, there could be no 
horizon, the extremal horizon or both of the Reissner-Nordstr\"om horizons 
depending on the relative values of $Q$ and $M$. There surely are unstable 
static solutions for all $\alpha$ and $\gamma^2$ but we do not consider those 
as a static solution. The radial component of normal is positive in all cases.
\midinsert{
\centerline{\epsfbox{static.eps}}
{\qquad\qquad De Sitter region \qquad\qquad $Q^2 < M^2$ \ \qquad\qquad
\qquad $Q^2 = M^2$}
}
\endinsert

\eject

{
{\bf\expandafter\sl\subsec{Oscillatory Solutions}} 
Whenever we have a local minimum in 
the potential, we have oscillatory solutions in which the shell oscillates 
about the static radius. In all oscillatory solutions, the de Sitter part of 
the spacetime is the same; it does not contain horizon and the sign of the 
radial normal is positive. The radial normal of the Reissner-Nordstr\"om part 
is also positive. We have five types of paths. Three of them do not cross any 
horizon and the others cross the extremal or both of the blackhole horizons, 
suggesting the possibility of fomation of the regular blackholes.

\midinsert{
\centerline {\epsfbox{bhosc1.eps}\qquad\qquad\qquad\qquad\epsfbox{bhosc2.eps}}
\qquad\  $Q^2 < M^2$\qquad\qquad\qquad\qquad\qquad\qquad\qquad\qquad\qquad\qquad $Q^2 < M^2$
}
\endinsert

\midinsert{
\centerline{\epsfbox{extosc1.eps}\qquad\qquad\qquad\qquad\epsfbox{extosc2.eps}}
\qquad\  $Q^2 = M^2$\qquad\qquad\qquad\qquad\qquad\qquad\qquad\qquad\qquad\qquad $Q^2 = M^2$
}
\endinsert
}

\midinsert{
\centerline {\epsfbox{nhosc.eps}}
\centerline{$Q^2 > M^2$}
}
\endinsert

{\bf\expandafter\sl\subsec{ Bounce and Monotonic Solutions}}
In these cases, the shell 
radius goes from zero or a minimum value to the infinity. From equation for 
the normal, we can see that for the large value of the radius, the radial 
component of normal outside the shell is negative. So the shell is not 
observable in the assymtotically flat region, $I_{+}$. But all of this case 
contains the de Sitter horizon inside the shell indicating the topological 
inflation.

\midinsert{
\centerline {\epsfbox{bhbounce11.eps}\qquad\qquad\qquad\qquad\epsfbox{bhbounce12.eps}}
\quad $Q^2 < M^2$(bounce)\qquad\qquad\qquad\qquad\qquad\qquad\qquad\qquad $Q^2 < M^2$(bounce)
}
\endinsert
\vfill
\eject

\vbox
{\midinsert{
\centerline{\epsfbox{bhbounce13.eps}\qquad\qquad\qquad\qquad\epsfbox{bhbounce21.eps}}
\quad $Q^2 < M^2$(bounce)\qquad\qquad\qquad\qquad\qquad\qquad\qquad\qquad$Q^2 < M^2$(bounce)
}
\endinsert
\nobreak
\midinsert{
\centerline {\epsfbox{bhbounce22.eps}\qquad\qquad\qquad\qquad\epsfbox{bhbounce23.eps}}
\quad$Q^2 < M^2$(bounce)\qquad\qquad\qquad\qquad\qquad\qquad\qquad\qquad$Q^2 < M^2$(bounce)
}
\endinsert
\nobreak
\midinsert{
\centerline{\epsfbox{bhbounce31.eps}\qquad\qquad\qquad\qquad\epsfbox{bhbounce32.eps}}
\quad$Q^2 < M^2$(bounce)\qquad\qquad\qquad\qquad\qquad\qquad\qquad\qquad$Q^2 < M^2$(bounce)
}
\endinsert
}
\vfill
\eject

{
{
\midinsert{
\centerline {\epsfbox{bhbounce33.eps}\qquad\qquad\qquad\qquad\epsfbox{bhmono1.eps}}
\quad$Q^2 < M^2$(bounce)\qquad\qquad\qquad\qquad\qquad\qquad\qquad$Q^2 < M^2$(monotonic)
}
\endinsert
\nobreak
\midinsert{
\centerline{\epsfbox{bhmono2.eps}\qquad\qquad\qquad\qquad\epsfbox{bhmono3.eps}}
\quad $Q^2 < M^2$(monotonic)\qquad\qquad\qquad\qquad\qquad\qquad\qquad$Q^2 < M^2$(monotonic)
}
\endinsert
}
}

{
\newsec {conclusion}
As we see in the previous sections, there are lots of possibilities in our 
solutions. We, however, concentrate on the solutions which are interesting such 
as regular black holes, ones which pass through the assymptotically flat region 
and soultions which contain De Sitter horizon inside the shell indicating the
possibility of topological inflation.

{\bf\expandafter{\sl\subsec{ Regular black holes}}}

Before we go further, we think how we can avoid the singularity. Arvin Bord 
showed that Bardeen black hole can avoid singularities by topology change from
open universe to closed universe. Our model in the non extremal case can be 
considered as a kind of Bardeen black hole. For the extremal case, there is no
change in topology. But we still have a regular black hole solution. We interpret
it as the following. According to Penrose, a singularity must occur after a 
trapped surface has formed. But in the extremal case in our model, there is no
trapped surface at all. So we could avoid singularity in extremal case, also.

{\bf{bounded solution}}- In the entire parameter space, we have bounded
black holes of which both normals are negative. We should think of the issue of
sign of the normals more deeply. If the normal is negative in $\threem$ region as
\boulware\ said, then we have two kind of bounded regular black holes; one is non
extremal and the other one is extremal. The solution of the case in which $Q^2 >
M^2$ is not the solution which satisfies the boundary condition. These bounded 
monopole black holes are not observable to the observer in the assymptotically
flat region.

{\bf{static solution}}- We have stable static solutions in some 
parameter space; for all $\gamma^2$ when $\alpha < \left(1 \over 2\right)^{4/3}$
and for $\gamma^2 < \gammamax$ when $\left(1 \over 2\right)^{4/3} < \alpha <
\left(3 \over 4^{4/3}\right)$. They are not observable, either. We should 
think about the perturbation (fluctuation in $E$ maintainning $\alpha$ and
$\gamma^2$ the same or decreasing $M$, due to black hole radiation?). We should 
study the relation between no-shell matching solution and this static solution.
In all paramater space, we have at least one unstable static solution. We,
however, don't consider those cases.

{\bf{oscillating solutions}}- The oscillating regular black holes are 
not observable. In other words, they don't pass through the assymptotically flat
region.

{\bf\expandafter{\sl\subsec{ Observable case}}} 

When parameter $\alpha$ is less than the certain value, $\left( 1 \over 2 \right
)^{4/3}$, we alway have observable black holes. These can be non extremal or
extremal depending on the relative values of $Q^2$ and $M^2$. They look like
oscillating in the $V-z$ plot. For a observer in the assymptotically flat region,
they will suddenly appear with some radius of either $\R+$ or $\Rext$ and grow to 
their maximum radius and decrease to the radius of their first appearance and 
then at at the radius, they will suddenly disappear beyond the black hole 
horizons.

{\bf\expandafter{\sl\subsec{ De Sitter horizon containing shell}}}

For the entire parameter range. we have this kind of solutions. All monotonic
solutions and bounce solutions have De Sitter horizon in the shell. But none of 
them are observable in the assymptotically flat region.

\newsec{things to study more}

We do not quite under stand the relation of sign of the radial component of outside normal and the spacetime diagram.
The relation between the static stable solution of no-shell situation and of 
with-shell situation should be studied more. I.e., the role of the surface 
energy density as a purterbation is not clear, yet.
}

\input conclusion.tex
\listrefs
\end


\vfill\eject
\midinsert{
\centerline{\epsfbox{dsfig21.eps}}
\centerline{\it Figure 1. \quad $V-\vds$ configuration $\I$.}
}
\endinsert\vskip 7mm
\midinsert{
\centerline{\epsfbox{dsfig22.eps}}
\centerline{\it Figure 2. \quad $V-\vds$ configuration $\II$.}
}
\endinsert\vskip 7mm
\vfill\eject
\midinsert{
\centerline{\epsfbox{dsfig23.eps}}
\centerline{\it Figure 3. \quad $V-\vds$ configuration $\III$.}
}
\endinsert\vskip 7mm
\midinsert{
\centerline{\epsfbox{dsfig24.eps}}
\centerline{\it Figure 4. \quad $V-\vds$ configuration $\IV$.}
}
\endinsert\vskip7mm
\vfill\eject
\midinsert{
\centerline{\epsfbox{dsfig25.eps}}
\centerline{\it Figure 5. \quad $V-\vds$ configuration $\V$.}
}
\endinsert




%
%

%

To make dimensionless variables, we used ratio of each parameter $Q$, $M$ and
$\sigma$ with $H$. That is the reason why we have three dimensionless variables
out of 4 parameters with dimensions. The cosmological constant, $H$, itself 
determines the relations between $z(t)$ and $\rshell$ and between $\tau'$ and
$\tau$. In \bgg, the potential is independent of one parameter, $M$ and they 
can plot one potential with different energies which are dependent on $M$ by
varying the mass. In our case, by fixing $\alpha$ and $\gamma$ to get one
potential, the ratio of $\left(QH \over (2MH)^{2/3}\right)$ is fixedx.
And wihout changing $\alpha$ and $\gamma$, i.e.,with the same potential, 
we can change the absolute value of $QH$ to have different energies.


\eqn\variables{\eqalign{
E &= {-(8 \pi \sigma)^2 \over {(2M)^{2/3}(H^2 + 16 {\pi}^2 {\sigma}^2)^{4/3}}} 
= - { \sqrt{\alpha} {\gamma}^2 (4-{\gamma}^2)^{1/2} \over {2(QH)}}\cr
\alpha &= {{Q^2 (H^2 + 16 {\pi}^2 {\sigma}^2 )^{1/3}} \over (2M)^{4/3}},\qquad 
{\gamma} = {8 \pi \sigma \over (H^2 + 16 {\pi}^2 {\sigma}^2)^{1/2}} 
\cr
}}

\vskip 0.1in\noindent
{\bf\it Other possibly useful formulae}

\eqn\normc{
\nout = {1\over{8 \pi \sigma \rshell}}{\left[-(H^2 + 16 \pi^2 \sigma^2)
\rshell^2 + {2M \over \rshell} - {Q^2 \over \rshell^2} \right]}
}
We express the equation \normc in terms of dimensionless variables and
plug it in the equation \norma to get $\nin$ as the following.

To understand the relationship between the 
locations of horizons and the shell, as Blau et. al. did in \bgg, we invert 
the above equation to get

%


{\bf\expandafter{ \sl\subsec{Possible configurations of potential and the 
Reissner-Nordstr\"om horizons}}}
{Compared with the potential- de Sitter horizon configuration, potential-
Reissner-Nordstr\"om horizon configuration shows complicated features 
depending on when the local minimum occurs; it occurs in one of three cases,
$Q < M$, $Q = M$ or $Q > M$. The followings are those possible configurations.}
\vskip2mm
\midinsert{
\centerline{\epsfbox{rnfig21.eps}}
\centerline{\it Figure 6. \quad $V-\vrn$ configuration $1$.}
}
\endinsert
\midinsert{
\centerline{\epsfbox{rnfig23.eps}}
\centerline{\it Figure 7. \quad $V-\vrn$ configuration $2$.}
}
\endinsert
\eject
\midinsert{
\centerline{\epsfbox{rnfig231.eps}}
\centerline{\it Figure 7a. \quad part of $V-\vrn$ configuration $2$.}
}
\endinsert
\midinsert{
\centerline{\epsfbox{rnfig24.eps}}
\centerline{\it Figure 8. \quad $V-\vrn$ configuration $3$.}
}
\endinsert
\eject
\midinsert{
\centerline{\epsfbox{rnfig241.eps}}
\centerline{\it Figure 8a. \quad part of $V-\vrn$ configuration $3$.}
}
\endinsert
\midinsert{
\centerline{\epsfbox{rnfig25.eps}}
\centerline{\it Figure 9. \quad $V-\vrn$ configuration $4$.} 
}
\endinsert
\eject
\midinsert{
\centerline{\epsfbox{rnfig251.eps}}
\centerline{\it Figure 9a. \quad part of $V-\vrn$ configuration $4$.}
}
\endinsert\vskip5mm
\midinsert{
\centerline{\epsfbox{rnfig26.eps}}
\centerline{\it Figure 10. \quad $V-\vrn$ configuration $5$.}
}
\endinsert
\eject
\midinsert{
\centerline{\epsfbox{rnfig261.eps}}
\centerline{\it Figure 10a. \quad part of $V-\vrn$ configuration $5$.}
}
\endinsert
\midinsert{
\centerline{\epsfbox{rnfig28.eps}}
\centerline{\it Figure 11. \quad $V-\vrn$ configuration $6$.}
}
\endinsert
\eject
\midinsert{
\centerline{\epsfbox{rnfig29.eps}}
\centerline{\it Figure 12. \quad $V-\vrn$ configuration $7$.}
}
\endinsert
\midinsert{
\centerline{\epsfbox{rnfig210.eps}}
\centerline{\it Figure 13. \quad $V-\vrn$ configuration $8$.}
}
\endinsert

\noindent
The following is the character table of Reissner-Nordstr\"om part of each
path.

\centerline{\vbox{\offinterlineskip
\hrule
\halign{&\vrule#&
  \strut\quad\hfill#\quad\cr
height2pt&\omit&&\omit&&\omit&&\omit&&\omit&\cr
&path\hfill&&range of z\hfill&&Sign of $\nout$\hfill&& Q and M\hfill&&Horizon
Crossing&\cr
height2pt&\omit&&\omit&&\omit&&\omit&&\omit&\cr
\noalign{\hrule}
height2pt&\omit&&\omit&&\omit&&\omit&&\omit&\cr
&a&&$0 < z \le \zmax$&&$-$&&$Q > M$&&*&\cr
&b&&$0 < z \le \zmax$&&$-$&&$Q = M$&&no&\cr
&c&&$0 < z \le \zmax$&&$-$&&$Q < M$&&no&\cr
height2pt&\omit&&\omit&&\omit&&\omit&&\omit&\cr
\noalign{\hrule}
height2pt&\omit&&\omit&&\omit&&\omit&&\omit&\cr
&st1&&$z = \zst$&&$+$&&$Q > M$&&*&\cr
&st2&&$z = \zst$&&$+$&&$Q = M$&&no&\cr
&st3&&$z = \zst$&&$+$&&$Q < M$&&no&\cr
height2pt&\omit&&\omit&&\omit&&\omit&&\omit&\cr
\noalign{\hrule}
height2pt&\omit&&\omit&&\omit&&\omit&&\omit&\cr
&d&&$\zmin \le z \le \zmax$&&$+$&&$ Q > M$&&*&\cr
&e&&$\zmin \le z \le \zmax$&&$+$&&$ Q = M$&&yes&\cr
&f&&$\zmin \le z \le \zmax$&&$+$&&$ Q = M$&&no&\cr
&g&&$\zmin \le z \le \zmax$&&$+$&&$ Q < M$&&yes&\cr
&h&&$\zmin \le z \le \zmax$&&$+$&&$ Q < M$&&no&\cr
height2pt&\omit&&\omit&&\omit&&\omit&&\omit&\cr
\noalign{\hrule}
height2pt&\omit&&\omit&&\omit&&\omit&&\omit&\cr
&i&&$\zmin \le z < \infty$&&$-$&&$ Q > M$&&*&\cr
&j&&$\zmin \le z < \infty$&&$-$&&$Q = M$&&**&\cr
&k&&$\zmin \le z < \infty$&&$-$&&$Q = M$&&yes&\cr
&l&&$\zmin \le z < \infty$&&$+ -$&&$Q = M$&&yes&\cr
&m&&$\zmin \le z < \infty$&&$-$&&$Q < M$&&**&\cr
&n&&$\zmin \le z < \infty$&&$-$&&$Q < M$&&yes&\cr
&o&&$\zmin \le z < \infty$&&$+ -$&&$Q < M$&&yes&\cr
&p&&$\zmin \le z < \infty$&&$- + -$&&$Q < M$&&yes&\cr
height2pt&\omit&&\omit&&\omit&&\omit&&\omit&\cr
\noalign{\hrule}
height2pt&\omit&&\omit&&\omit&&\omit&&\omit&\cr
&q&&$0 < z < \infty$&&$- $&&$Q > M$&&*&\cr
&r&&$0 < z < \infty$&&$-$&&$Q = M$&&yes&\cr
&s&&$0 < z < \infty$&&$- + -$&&$ Q < M$&&yes&\cr
&t&&$0 < z < \infty$&&$-$&&$Q < M$&&yes&\cr
height2pt&\omit&&\omit&&\omit&&\omit&&\omit&\cr}
\hrule}}
\centerline{\it Table 2. Character of each path of the shell-
Reissner-Nordstr\"om part}
\midinsert\narrower\narrower{
{\ninerm *:There is no horizon because $Q > M$.\hfill

**:There is no horizon in the spacetime since they(it) occur(s) inside the
   shell.}
}
\endinsert\vskip5mm